\documentclass[a4paper,11pt]{article}
\pdfoutput=1 

\usepackage[utf8]{inputenc}

\usepackage{jcappub} 

\usepackage[T1]{fontenc} 

\usepackage[utf8]{inputenc}
\usepackage{siunitx}
\usepackage{url}
\usepackage{tensor}
\usepackage{longtable}
\usepackage{float}
\usepackage{mathtools}
\usepackage{amssymb}
\usepackage{amsmath}
\usepackage{makecell,tabularx,multirow}
\usepackage{color, colortbl}
\usepackage{adjustbox}
\usepackage{tikz}
\usepackage[pages=some]{background}
\usepackage{hyperref}
\usepackage{accents}
\usepackage{todonotes}
\usepackage{eqnarray}
\usepackage{makecell}
\usepackage[margin=1in]{geometry}
\hypersetup{
    colorlinks=true,
    linkcolor=blue,
    filecolor=magenta,
    citecolor=red
}
\usepackage{comment}
\usepackage{rotating}

\newcolumntype{C}{>{\centering\arraybackslash}X}
\newcolumntype{x}[1]{>{\centering\arraybackslash\hspace{0pt}}p{#1}}

\newcommand{\udt}[3]{#1^{#2}_{\phantom{#2}#3}}
\newcommand{\udut}[4]{#1^{#2\phantom{#3}#4}_{\phantom{#2}#3\phantom{#4}}}

\newcommand{\dut}[3]{#1_{#2}^{\phantom{#2}#3}}
\newcommand{\dudt}[4]{#1_{#2\phantom{#3}#4}^{\phantom{#2}#3}}

\newcommand{\lc}[1]{\accentset{\circ}{#1}}

\title{\boldmath Ghost and Laplacian Instabilities in  Teleparallel  Horndeski Gravity}

\author[a,b,c]{Salvatore Capozziello,} 
\author[d,e,1]{Maria Caruana,\note{Corresponding author}}
\author[d,e]{Jackson Levi Said,}
\author[f]{Joseph Sultana}

\affiliation[a]{Dipartimento di Fisica "E. Pancini", Universit\`a degli Studi di Napoli, "Federico II",  Complesso Universitario Monte S. Angelo, Via Cinthia 9 Edificio G, 80126 Napoli, Italy.}
\affiliation[b]{Istituto Nazionale di Fisica Nucleare (INFN), Sezione di Napoli Complesso Universitario Monte S. Angelo, Via Cinthia 9 Edificio G, 80126 Napoli, Italy.}
\affiliation[c]{Scuola Superiore Meridionale, Largo San Marcellino 10, 80138 Napoli, Italy.}
\affiliation[d]{Institute of Space Sciences and Astronomy, University of Malta, Malta, MSD 2080.}
\affiliation[e]{Department of Physics, University of Malta, Malta, MSD 2080.}
\affiliation[f]{Department of Mathematics, University of Malta, Msida, Malta.}

\emailAdd{capozziello@na.infn.it}
\emailAdd{maria.caruana.16@um.edu.mt}
\emailAdd{jackson.said@um.edu.mt}
\emailAdd{joseph.sultana@um.edu.mt}

\abstract{Teleparallel geometry offers a platform on which to build up theories of gravity where torsion rather than curvature mediates gravitational interaction. The teleparallel analogue of Horndeski gravity is an approach to teleparallel geometry where scalar-tensor theories are considered in this torsional framework. Teleparallel gravity is based on the tetrad formalism. This turns out to result in a more general formalism of Horndeski gravity. In other words, the class of teleparallel Horndeski gravity models is much broader than the standard metric one. In this work, we explore constraints on this wide range of  models coming from ghost and Laplacian instabilities. The aim is   to limit  pathological branches of the theory by fundamental considerations. It is possible to conclude that a  very large class of models  results physically viable.}

\keywords{cosmology, cosmological perturbation, scalar-tensor gravity, teleparallel gravity}

\begin{document}
\maketitle
\flushbottom

\section{Introduction} \label{sec:intro}

General relativity (GR) has been the gravitational foundation of cosmology for over a century. Its latest embodiment appears in a  picture  wherein the evolutionary processes of the Universe are described in the framework of the so-called  $\Lambda$CDM model \cite{misner1973gravitation,Clifton:2011jh}. Supported by overwhelming observational evidences, this model describes a Universe  started with a big bang, then driven through an inflationary phase and other well known epochs to eventually produce an accelerating Universe at late-times \cite{Riess:1998cb,Perlmutter:1998np}. In the $\Lambda$CDM model, the late time acceleration expansion is driven by the cosmological constant or some form of dark energy. Despite a lot of foundational works, internal consistency issues persist in this respect \cite{RevModPhys.61.1,Appleby:2018yci,Ishak:2018his}. Recently, observational challenges to the standard model has arisen in the form of cosmological tensions with  statistically significant differences between predictions of  expansion  from early time data \cite{Bernal:2016gxb,DiValentino:2020zio,DiValentino:2021izs} and measurements from  late time  \cite{Riess:2019cxk,Wong:2019kwg}. These tensions continue to increase with new survey data \cite{Riess:2021jrx,Brout:2021mpj,Scolnic:2021amr}, and may permeate into other sectors of cosmology besides expansion \cite{Abdalla:2022yfr,DiValentino:2020vvd, Capozziello:2020nyq}. This situation  leads to take into account   potential alternatives to the standard cosmological $\Lambda$CDM model   in the context of possible modifications to the gravitational sector.

A solution to the problem of cosmic tensions, as well as of other longstanding issues,  can be offered by alternative theories of gravity, in particular by scalar-tensor gravity. Scalar fields improve GR in view of Mach principle  and could naturally address several  issues   in cosmology and astrophysics like inflation, dark matter and dark energy \cite{Book}. The most general scalar-tensor theory of gravity, where scalar fields are minimally and non-minimally coupled to curvature,  is the so-called Horndeski gravity \cite{Horndeski:1974wa}. It is the most general theory of gravity  producing second order field equations. This is advantageous because nature appears to produce, generally, second order equations of motion. Furthermore, higher order theories can  produce Ostrogradsky instabilities making second order theories more attractive from a fundamental perspective \cite{Woodard:2006nt,Motohashi:2014opa,Pagani:1987ue}. Horndeski gravity is formulated by four free functions of the scalar field and its kinetic term \cite{Miranda:2022brj,Miranda:2022wkz}. This offers a rich phenomenology on which to produce  dark energy models but also dark matter and inflation. However, recent multimessenger observations by the LIGO collaboration, in the gravitational event GW170817 \cite{TheLIGOScientific:2017qsa}, and measurements of the companion electromagnetic counterpart, namely GRB170817A \cite{Goldstein:2017mmi}, has placed stringent constraints on the speed of propagation of gravitational waves (GW) to within deviations of at most one part in $10^{15}$. Considering these constraints, many  branches of Horndeski gravity have been disqualified in regular curvature-based gravity \cite{Kobayashi:2011nu,Gleyzes:2013ooa,Koyama:2015vza}.
However, Horndeski gravity has a number of interesting generalizations  \cite{Kobayashi:2019hrl} where higher order terms, avoiding the Ostrogradsky instabilities, can be incorporated into the theory. Another possibility is to reconsider the geometric foundations of curvature on which  Horndeski gravity is built. Curvature can be  expressed through the Levi-Civita connection ${\lc{\Gamma}}^{\lambda}_{\mu\nu}$ 
obtained from the metric \cite{misner1973gravitation}. Here an over-circle represents quantities determined by  the Levi-Civita connection.
 
In teleparallel gravity (TG), the teleparallel connection (${\Gamma}^{\lambda}_{\mu\nu}$) replaces the Levi-Civita connection and so dynamics of gravity is replaced from curvature to torsion \cite{Bahamonde:2021gfp,Aldrovandi:2013wha,Cai:2015emx,Krssak:2018ywd}. The teleparallel connection is curvatureless  and satisfies metricity. For this reason, the teleparallel Ricci scalar turns out to identically vanish, i.e. $R=0$ (this is not to say that the standard Ricci scalar is zero, being, in general,  $\lc{R} \neq0$). On the other hand, TG naturally produces a torsion scalar ($T$) \cite{Aldrovandi:2013wha}, which  turns out to be equal to the curvature-based Ricci scalar up to a boundary term ($B$). For this reason, the linear torsion scalar produces the \textit{Teleparallel equivalent of General Relativity} (TEGR) which is dynamically equivalent to GR in the classical regime. This distinction  between the torsion scalar,  producing second order terms in the equations of motion, and the boundary term,  producing fourth order terms in the equations of motion (when the boundary term is nonlinear in the action), gives rise to a much richer landscape of theories on which to build cosmological models, if compared with the standard Einstein-Hilbert action. 

Another way to conceptualize this feature is through the teleparallel analogue of the Lovelock theorem \cite{Lovelock:1971yv,Gonzalez:2015sha,Bahamonde:2019shr} which produces a much wider range of actions with second order field equations. Equivalences and differences of these representations are discussed in Ref. \cite{Capozziello:2022zzh}.

Similar to GR, TEGR features many modifications such as $f(T)$ gravity \cite{Ferraro:2006jd,Ferraro:2008ey,Bengochea:2008gz,Linder:2010py,Chen:2010va,Bahamonde:2019zea, RezaeiAkbarieh:2018ijw,Cai:2015emx,Farrugia:2016pjh} which is generically second order in nature, as well as $f(T,B)$ gravity \cite{Bahamonde:2015zma,Bahamonde:2016grb,Paliathanasis:2017flf,Farrugia:2018gyz,Bahamonde:2016cul,Wright:2016ayu,Farrugia:2020fcu,Capozziello:2019msc,Farrugia:2018gyz,Escamilla-Rivera:2019ulu} which now features fourth order contributions similar to $f(\lc{R})$ gravity \cite{Capozziello:2002rd, Sotiriou:2008rp,Faraoni:2008mf,Capozziello:2011et, Book,Nojiri:2006ri,Nojiri:2010wj,Nojiri:2017ncd}. There have also been various formulations of scalar--tensor gravity. See, e.g. Refs.\cite{Hohmann:2018vle,Hohmann:2018dqh,Hohmann:2018ijr}. 

Equipped with the TG formalism, one can consider the teleparallel analogue of Horndeski gravity \cite{Bahamonde:2019shr}. It is worth noticing that tetrads can be combined  in  roots of the metrics, as we  will discuss in the next section. Due to this feature, the TG analogue of Horndeski gravity turns out to be a much richer formalism. This feature  directly comes from the teleparallel analogue of the Lovelock theorem. This larger framework of models means that there are many more classes of theories that now satisfy the speed of GWs constraint \cite{Bahamonde:2019ipm,Bahamonde:2021dqn} as required from the above mentioned multimessenger analysis \cite{Ezquiaga:2017ekz}. This class of gravitational models also satisfies the parameterized post-Newtonian observational constraints \cite{Bahamonde:2020cfv} for a large number of models. The teleparallel analogue of Horndeski gravity also produces a number of interesting cosmologies such as those which are well-tempered \cite{Bernardo:2021izq,Bernardo:2021bsg}, as well as those that are derived from Noether symmetry considerations \cite{Capozziello:2016eaz,Capozziello:2018gms,Dialektopoulos:2018qoe, Dialektopoulos:2021ryi, Bajardi:2022ypn}. In this way, all models that were previously disqualified in regular Horndeski may be revived in this new formalism based on TG.

In this work, we perform a stability analysis of the teleparallel analogue of Horndeski gravity through considerations on a Minkowski spacetime background. The Minkowski background is an ideal arena to probe the fundamental behavior of theories with large classes of models since any astrophysical or cosmological setting must first be stable on a Minkowski spacetime. In our analysis, we consider both background and leading order perturbations in the context of a scalar-vector-tensor decomposition. This is important to fully decompose any potentially problematic part of the theory in detail. In the TG setting, this is more intricate since the metric tensor ($g_{\mu\nu}$) is replaced as the fundamental dynamical variable of the field equations with the tetrad ($\udt{e}{A}{\mu}$) and the spin connection ($\udt{\omega}{A}{B\mu}$), which are discussed in detail later on. In all cases, we focus our study on the ghost and gradient instabilities, which can wreak havoc on gravitational models since this may, respectively, produce scalar field kinetic terms with the wrong sign and unbounded propagation speeds  of particular perturbations, which can lead to unphysical models. 

The structure of the manuscript is the following: in Sec.~\ref{sec:BDLS_intro}, we summarize TG and its teleparallel analogue of Horndeski gravity; this then leads to the Minkowski background equations of motion in Sec.~\ref{sec:min_bg_eom}. Scalar--vector--tensor perturbations are discussed in  Sec.~\ref{sec:perturbations}.  From these perturbations, we are able to explore potential instabilities in Sec.~\ref{sec:instabilities} where our main results are reported.  Discussion and conclusions are drawn  in Sec.~\ref{sec:conclu}. In this work, we use geometric units and the signature $(-,+,+,+)$.

\section{Teleparallel  Horndeski Gravity} \label{sec:BDLS_intro}

GR, a curvature-based theory, is constructed on torsionless Levi-Civita connection $\lc{\Gamma}^{\lambda}_{\mu\nu}$, where, as said above,  the overhead circle (\,$\lc{}$\,) represents quantities built with this connection. This leads to numerous theories of gravity beyond GR as the gravitational field can be expressed through the Riemann tensor and its contractions such as the Ricci tensor and the Riemann scalar \cite{Capozziello:2011et}, the latter produces the Einstein-Hilbert action \cite{misner1973gravitation}. On the other hand, TG incorporates the teleparallel connection, dubbed as teleparallel connection $\Gamma^{\lambda}_{\mu\nu}$, leading to a theory satisfying the curvature-less and metricity conditions \cite{Krssak:2018ywd,Aldrovandi:2013wha,Cai:2015emx,Bahamonde:2021gfp}, resulting in a torsionful theory. Later in this section, it will be shown that even though the Ricci scalar $R$ vanishes, due to the symmetry of the connection, the curvature-ful connection gives a non-zero value for the Riemann tensor. Hence, the Ricci scalar $\lc{R}$ can be defined through teleparallel quantities.

The fundamental dynamical object of GR is the metric tensor $g_{\mu\nu}$, but within TG, the metric is expressed through the tetrad $\udt{e}{A}{\mu}$, which acts as a transformation between the local and general manifold spaces. The tetrad and the inertial spin connection $\udt{\omega}{B}{C\nu}$ become the fundamental objects~\cite{Aldrovandi:2013wha} while creating a link between the general manifold denoted through Greek indices to the local Minkowski manifold denoted by Latin indices. Thus, the relationship between Minkowski and general spacetimes is given by~\cite{Hehl:1976kj}
\begin{align}\label{metric-definition}
    g_{\mu\nu} = \eta_{AB}\,\udt{e}{A}{\mu}\,\udt{e}{B}{\nu}\,,
    \qquad \qquad\qquad
    \eta_{AB} = g_{\mu\nu}\,\dut{E}{A}{\mu}\,\dut{E}{B}{\nu}\,,
\end{align}
where $\dut{E}{A}{\mu}$ is the inverse tetrad which satisfies orthogonality conditions
\begin{align}
    \udt{e}{A}{\mu}\,\dut{E}{B}{\mu} = \delta^{A}_{B}\,,
    \qquad \qquad\qquad\qquad
    \udt{e}{A}{\mu}\, \dut{E}{A}{\nu} = \delta^{\nu}_{\mu}\,.
\end{align}
The teleparallel connection is defined through the TG variables as~\cite{Cai:2015emx,Krssak:2018ywd} 
\begin{align}
    \Gamma^{\lambda}_{\mu\nu} = \dut{E}{A}{\lambda} \left( \partial_{\nu} \udt{e}{A}{\mu} + \udt{\omega}{A}{B\nu} \udt{e}{B}{\mu} \right)\,.
\end{align}
Note, the quantities without an overhead circle will correspond to those objects that are related to teleparallel gravity or calculated on its connection. The condition
\begin{align}
    \partial_{[\mu}\udt{\omega}{A}{|B|\nu]} + \udt{\omega}{A}{C[\mu}\udt{\omega}{C}{|B|\nu]} \equiv 0 \,,
\end{align}
results in a flat spin connection~\cite{Bahamonde:2021gfp}, where square brackets denote the antisymmetric operator, and can equivalently be used to determine the components of the spin connection. The local Lorentz transformation (LLT), wherein $\udt{\Lambda}{A}{B}$ represents Lorentz boosts and rotations, is used to define the spin connection $\udt{\omega}{A}{B\mu} = \udt{\Lambda}{A}{C}\,\partial_{\mu}\dut{\Lambda}{B}{C}$~\cite{Aldrovandi:2013wha}. It plays a role in the field equations since an infinite number of tetrad choices could satisfy Eq.~\eqref{metric-definition} for a particular metric, thus the spin connection is used to counterbalance inertial effects. This ensures the theory, in this case TG, remains covariant~\cite{Krssak:2015oua}. Moreover, there exists a Lorentz frame such that the spin connection is set to zero. It is referred to as the Weitzenb\"{o}ck gauge~\cite{Weitzenbock:1923efa,Krssak:2018ywd}. Hence, from this point onwards, the spin connection will be dropped following the application of the aforementioned gauge.

Similar to how the Riemann tensor, constructed on the Levi-Civita connection, is associated to the curvature property of GR, the analogy of this for TG is the torsion tensor defined by the teleparallel connection~\cite{Aldrovandi:2004db,Krssak:2018ywd}
\begin{align}
    \udt{T}{A}{\mu\nu} = \Gamma^{A}_{\nu\mu} - \Gamma^{A}_{\mu\nu}\,. 
\end{align}
While curvature in GR is defined as a deformation of spacetime, the torsion tensor in TG represents the field strength of gravitation that transforms covariantly under LLTs and diffeomorphisms~\cite{Krssak:2015oua}. Another important aspect of TG is that the torsion tensor can be decomposed in irreducible parts~\cite{Hayashi:1979qx, Capozziello:2001mq, Bahamonde:2017wwk}, namely
\begin{align}
    a_{\mu} &= \frac{1}{6} \epsilon_{\mu\nu\lambda\rho}\,T^{\nu\lambda\rho}\,,\\
    v_{\mu} &= \udt{T}{\lambda}{\lambda\nu}\,,\\
    t_{\lambda\mu\nu} &= \frac{1}{2}(T_{\lambda\mu\nu} + T_{\mu\lambda\nu}) + \frac{1}{6}(g_{\nu\lambda}v_{\mu} + g_{\nu\mu}v_{\lambda}) - \frac{1}{3}g_{\lambda\mu} v_{\nu}\,,
\end{align}
giving the axial, vectorial and tensorial pieces, respectively, and $\epsilon_{ABCD}$ being the four-dimensional Levi-Civita tensor. Hence, the respective scalar invariants are given by~\cite{Capozziello:2001mq,Bahamonde:2015zma}
\begin{align}
    T_{\text{ax}} &= a_{\mu}a^{\mu} = -\frac{1}{18} T_{\lambda\mu\nu}(T^{\lambda\mu\nu} - 2T^{\mu\lambda\nu})\,,\\
    T_{\text{vec}} &= v_{\mu}v^{\mu} = \udt{T}{\lambda}{\lambda\mu} \dut{T}{\rho}{\rho\mu}\,,\\
    T_{\text{ten}} &= t_{\lambda\mu\nu} t^{\lambda\mu\nu} = \frac{1}{2}T_{\lambda\mu\nu}(T^{\lambda\mu\nu} + T^{\mu\lambda\nu}) - \frac{1}{2}\udt{T}{\lambda}{\lambda\mu} \dut{T}{\rho}{\rho\mu}\,,
\end{align}
which, under parity transformation, are invariant  scalars. On the other hand, terms such as $P_{1} = v^{\mu} a_{\mu}$ and $P_{2} = \epsilon_{\mu\nu\sigma\rho}t^{\lambda\mu\nu}\udt{t}{\lambda}{\rho\sigma}$ are excluded due to parity-violation~\cite{Hayashi:1979qx,Bahamonde:2019shr}. The combination of the scalar invariants leads to the torsion scalar
\begin{align}\label{Torsion-Scalar}
    T = \frac{3}{2}T_{\text{ax}} + \frac{2}{3}T_{\text{ten}} - \frac{2}{3}T_{\text{vec}} = \frac{1}{2}\left(\dut{E}{A}{\lambda} g^{\rho\mu} \dut{E}{B}{\nu} - 2 \dut{E}{B}{\rho} g^{\lambda\mu} \dut{E}{A}{\nu} + \frac{1}{2}\eta_{AB} g^{\mu\rho} g^{\nu\lambda}\right) \udt{T}{A}{\mu\nu} \udt{T}{B}{\rho\lambda}\,.
\end{align}
An identical result can be obtained through contraction between the torsion tensor and the superpotential $\dut{S}{A}{\mu\nu}$ which represents the potential relation of the gravitational energy-momentum tensor~\cite{Nesseris:2013jea,Aldrovandi:2004db,Koivisto:2019ggr}:
\begin{align}
    \dut{S}{A}{\mu\nu} = \frac{1}{2} \left(\udt{K}{\mu\nu}{A} - \dut{E}{A}{\nu} \udt{T}{\alpha\mu}{\alpha} + \dut{E}{A}{\mu} \udt{T}{\alpha\nu}{\alpha}\right)\,,
\end{align}
where
\begin{align}
    \udt{K}{\lambda}{\mu\nu} = \Gamma^{\lambda}_{\mu\nu} - \lc{\Gamma}^{\lambda}_{\mu\nu} = \frac{1}{2}\left(\dudt{T}{\mu}{\lambda}{\nu} + \dudt{T}{\nu}{\lambda}{\mu} - \udt{T}{\lambda}{\mu\nu}\right)
\end{align}
is the contorsion tensor relating TG and GR. These quantities can be used to form the torsion scalar~\cite{Cai:2015emx}, written as
\begin{align}
    T = \dut{S}{A}{\mu\nu}\udt{T}{A}{\mu\nu}\,,
\end{align}
which can be shown to be equivalent to the result obtained in Eq.~\eqref{Torsion-Scalar}. The Ricci scalar dependent on the teleparallel connection, which vanishes, can be related, using the contorsion tensor, to the regular Ricci scalar. This results in a relationship between the Levi-Civita and the teleparallel connections defined Ricci scalar~\cite{Hehl:1976kj,Bahamonde:2015zma}
\begin{align}
    R = \lc{R} + T - B = 0\,,
\end{align}
where $B = \tfrac{2}{e} \partial_{\mu}(e\,\udut{T}{\lambda}{\lambda}{\mu}) = 2 \lc{\nabla}_{\mu}\udut{T}{\lambda}{\lambda}{\mu}$ is a boundary term and  $e = \text{det}(\udt{e}{A}{\mu}) = \sqrt{-g}$ is the tetrad determinant. Hence, the curvature-ful Ricci scalar is  non-vanishing being
\begin{align}
    \lc{R} = -T + B\,.
\end{align}
The total divergence term found in $B$ accounts for the fourth order derivative contributions to the field equations in modified theories of gravity~\cite{Bahamonde:2015zma,Capozziello:2019msc}. This is embodied within the Ricci scalar in GR~\cite{Sotiriou:2008rp,Capozziello:2011et}. Thus, the Teleparallel  Equivalent of General Relativity (TEGR)~\cite{Hehl:1994ue,BeltranJimenez:2019esp} is defined as the linear appearance of the torsion scalar since both Lagrangians are equal up to a boundary term.

The equivalence principle in GR allows one to raise local Lorentz frames from a Minkowski metric to a general metric tensor, and additionally, partial derivatives are raised to covariant derivatives defined by the Levi-Civita connection~\cite{misner1973gravitation}. This procedure, referred to as the minimal coupling prescription, if applied within TG, is preserved for additional fields. Minkowski tetrads are raised to arbitrary ones and tangent space partial derivatives are exchanged for covariant derivatives based on Levi-Civita connection~\cite{Aldrovandi:2013wha,BeltranJimenez:2020sih}
\begin{align}
    \partial_{\mu} \rightarrow \lc{\nabla}_{\mu}\,.
\end{align}
Thus, by having both gravitational and scalar fields well developed, it is  possible to look at the teleparallel analogue of Horndeski gravity, referred to at times as Bahamonde-Dialektopoulos-Levi Said (BDLS) theory~\cite{Bahamonde:2019shr,Bahamonde:2019ipm,Bahamonde:2020cfv}. The construction of BDLS theory depends on the following criteria: (1) field equations are at most second order with respect to the tetrad and scalar; (2) as previously mentioned, scalar invariants do not violate parity; (3) contractions of the torsion tensor are at most quadratic~\cite{Bahamonde:2019shr}. All of these requirements allow for an adequate extension of the standard metric  Horndeski gravity. Note that Lovelock's theorem states that Einstein fields equations are the only second-order field equations from a Lagrangian density constructed through a four-dimensional metric. TG allows for the weakening of Lovelock~\cite{Lovelock:1971yv,Gonzalez:2015sha} theory as an additional scalar field $\phi$ is introduced, giving rise to further terms in the gravitational action.

Starting off  with the non-minimal coupling of scalar and torsion field, the linear contraction is given by~\cite{Bahamonde:2019shr}
\begin{align}
    I_{2} = v^{\mu} \lc{\nabla}_{\mu}\phi\,,
\end{align}
being the scalar $I_{1} = t^{\lambda\mu\nu} \lc{\nabla}_{\lambda}\phi\lc{\nabla}_{\mu}\phi\lc{\nabla}_{\mu}\phi = 0$, due to the complete symmetry of the tensor decomposition, Furthermore $I_{3} = a^{\lambda}\lc{\nabla}_{\lambda}\phi$ violates the parity condition due to an odd number of axial parts. Moreover, the quadratic contractions of this nature are given by~\cite{Bahamonde:2019shr}
\begin{align}
    J_{1} &= a^{\mu}a^{\nu}\lc{\nabla}_{\mu}\phi\lc{\nabla}_{\nu}\phi\,,\\
    J_{3} &= v_{\lambda} t^{\lambda\mu\nu}\lc{\nabla}_{\mu}\phi\lc{\nabla}_{\nu}\phi\,,\\
    J_{5} &= t^{\lambda\mu\nu} \dudt{t}{\lambda}{\alpha}{\nu}\lc{\nabla}_{\mu}\phi\lc{\nabla}_{\alpha}\phi\,,\\
    J_{6} &= t^{\lambda\mu\nu} \dut{t}{\lambda}{\alpha\beta}\lc{\nabla}_{\mu}\phi\lc{\nabla}_{\nu}\phi\lc{\nabla}_{\alpha}\phi\lc{\nabla}_{\beta}\phi\,,\\
    J_{8} &= t^{\lambda\mu\nu} \dut{t}{\lambda\mu}{\alpha}\lc{\nabla}_{\nu}\phi\lc{\nabla}_{\alpha}\phi\,,\\
    J_{10} &= \udt{\epsilon}{\mu}{\nu\lambda\rho} a^{\nu} t^{\alpha\rho\lambda}\lc{\nabla}_{\mu}\phi\lc{\nabla}_{\alpha}\phi\,,
\end{align}
while other possibilities are eliminated as they have already been included: 
\begin{align}
    J_{2} &= v^{\mu}v^{\nu}\lc{\nabla}_{\mu}\phi\lc{\nabla}_{\nu}\phi = I_{2}^{2}\,, \quad
    J_{4} = v_{\mu} t^{\lambda\mu\nu} \lc{\nabla}_{\lambda}\phi\lc{\nabla}_{\nu}\phi = J_{3}\,,\nonumber\\
    J_{7} &= t^{\lambda\mu\nu} \udt{t}{\alpha\beta}{\lambda}\lc{\nabla}_{\mu}\phi\lc{\nabla}_{\nu}\phi\lc{\nabla}_{\alpha}\phi\lc{\nabla}_{\beta}\phi = -2J_6\,,
\end{align}
while $J_{9} = t^{\lambda\mu\nu} t^{\alpha\beta\gamma}\lc{\nabla}_{\lambda}\phi\lc{\nabla}_{\mu}\phi\lc{\nabla}_{\nu}\phi\lc{\nabla}_{\alpha}\phi\lc{\nabla}_{\beta}\phi\lc{\nabla}_{\gamma}\phi = 0$ due to the total symmetry of the tensor irreducible part. Hence, BDLS action is described by~\cite{Bahamonde:2019shr}
\begin{align}\label{teledeski-action}
    \mathcal{S}_{\text{BDLS}} = \frac{1}{2\kappa^{2}} \int d^{4}x \, e\, \mathcal{L}_{\text{Tele}} + \frac{1}{2\kappa^{2}} \sum_{i=2}^{5} \int d^{4}x \, e \, \mathcal{L}_{i} + \int d^{4}x \, e \, \mathcal{L}_{\text{m}}\,,
\end{align}
where 
\begin{align}
    \mathcal{L}_{\text{Tele}} = G_{\text{Tele}}(\phi,X,T,T_{\text{ax}},T_{\text{vec}},I_{2},J_{1},J_{3}.J_{5},J_{6},J_{8},J_{10})\,.
\end{align}
Here $G_{\text{Tele}}$ is an arbitrary function, $X \coloneqq -\tfrac{1}{2} \partial^{\mu}\phi \partial_{\mu}\phi$ and
\begin{align}
    \mathcal{L}_{2} &\coloneqq G_{2}(\phi,X)\,,\\
    \mathcal{L}_{3} &\coloneqq -G_{3}(\phi,X)\lc{\Box}\phi\,,\\
    \mathcal{L}_{4} &\coloneqq G_{4}(\phi,X)(-T+B) + G_{4,X}(\phi,X)[(\lc{\Box}\phi)^{2} - \lc{\nabla}_{\mu}\lc{\nabla}_{\nu}\phi\,\lc{\nabla}^{\mu}\lc{\nabla}^{\nu}\phi]\,,\\
    \mathcal{L}_{5} &\coloneqq G_{5}(\phi,X) \lc{G}_{\mu\nu} \lc{\nabla}_{\mu}\lc{\nabla}_{\nu}\phi \nonumber \\ & \quad - \frac{1}{6} G_{5,X}(\phi,X)[(\lc{\Box}\phi)^{3} + 2 \lc{\nabla}_{\mu}\lc{\nabla}^{\nu}\phi\lc{\nabla}_{\nu}\lc{\nabla}^{\alpha}\phi\lc{\nabla}_{\alpha}\lc{\nabla}^{\mu}\phi - 3\lc{\nabla}_{\mu}\lc{\nabla}_{\nu}\phi\lc{\nabla}^{\mu}\lc{\nabla}^{\nu}\phi\,\lc{\square}\phi]\,,
\end{align}
which are identical to the standard metric Horndeski Lagrangians~\cite{Horndeski:1974wa} but calculated using the tetrad. Finally, $\mathcal{L}_{\text{m}}$ is the matter Lagrangian in Jordan conformal frame, $\lc{G}_{\mu\nu}$ is the Einstein tensor, and $\kappa^{2} = 8 \pi G$ where $G$ is the gravitational constant.

\section{Minkowski Background Equations in Teleparallel  Horndeski Gravity}\label{sec:min_bg_eom}

Let us explore now perturbations on a Minkowski background within the gravitational theory of teleparallel analogue of Horndeski. Firstly, we tackle the background equations for a general flat Friedman–Lema\^{i}tre\,–Robertson–Walker (FLRW) metric to obtain the necessary constraints. Then, this is followed by the analysis for perturbation theory up to second order which is needed for our approach since we consider the Euler-Lagrange equations to get the cosmological equations of motion.

The flat FLRW metric in Cartesian coordinates is given by
\begin{align}
    ds^{2} = -N(t)^{2} dt^{2} + a(t)^{2}(dx^{2} + dy^{2} + dx^{2})\,,
\end{align}
where $N(t)$ is the lapse function and $a(t)$ is the scale factor. As previously mentioned, this follows the metric signature that is mostly positive. The tetrad can be written in diagonal form as \cite{Krssak:2018ywd}
\begin{align}
    \udt{e}{A}{\mu} = \text{diag}(N(t),a(t),a(t),a(t))\,,  
\end{align}
which is compatible with the Weitzenb\"{o}ck gauge giving a vanishing spin connection. Hence, the action in Eq.~\eqref{teledeski-action} can be re-expressed in terms of these background quantities. Friedman equations are then obtained by varying with respect to the lapse function and scale factor. Additionally, the scalar field Klein-Gordon equation can be obtained by varying with respect to the scalar field~\cite{Bahamonde:2019shr}. Since we will be working in Minkowski background, the limits $N(t)\rightarrow 1$ and $a(t)\rightarrow 1$ are taken. Hence, the constraints obtained from the Friedman equation and scalar field variation in Minkowski background are given by
\begin{align}\label{background-equations-1}
   0 &= -G_{2} - G_{\text{Tele}} + 2 X G_{2,X} - 2X G_{3,\phi} + 2XG_{\text{Tele},X}\,,\\
   \label{background-equations-2}0 &= G_{2} + G_{\text{Tele}} - 2X G_{3,\phi} + 4X G_{4,\phi\phi} + 2 \ddot{\phi}G_{4,\phi}-2X \ddot{\phi} G_{3,X} + 4 X \ddot{\phi}G_{4,\phi X} - \tfrac{d}{dt}(\dot{\phi}G_{\text{Tele},I_{2}})\,,\nonumber\\
   & \\
   \label{background-equations-3}0 &= G_{2,\phi}+ G_{\text{Tele},\phi} - 2 X G_{2,\phi X} + 2X G_{3,\phi\phi} - 2X G_{\text{Tele},\phi X} - G_{2,X} \ddot{\phi} + 2G_{3,\phi}\ddot{\phi} - G_{\text{Tele},X} \ddot{\phi} \nonumber \\
   & \qquad - 2X \ddot{\phi}G_{\text{Tele},X} - 2X\ddot{\phi}G_{2,XX} + 2X \ddot{\phi}G_{3,\phi X} - 2X \ddot{\phi} G_{\text{Tele},XX}\,,
\end{align}
where all scalar invariants are background quantities and comma (\,,\,) denotes partial derivative. Moreover, the scalar field equation can also be expressed in terms of a Klein-Gordon equation as shown in Ref.~\cite{Bahamonde:2019shr}.

\section{Second Order Perturbed Action with Scalar-Vector-Tensor Decomposition}\label{sec:perturbations}

Here, we calculate the perturbed action up to second order terms which are necessary for  the Euler-Lagrange method to find the equations of motion of the system. By taking perturbations about a Minkowski background for both tetrad and scalar field, we can build the action up to second order. The tetrad and scalar perturbation are respectively given by
\begin{align}
    \udt{e}{A}{\mu} &\rightarrow \udt{e}{A}{\mu} + \epsilon\,\delta\udt{e}{A}{\mu} = \delta^{A}_{\mu} + \epsilon \, \udt{\delta e}{A}{\mu}\,,\\
    \phi &\rightarrow \phi + \epsilon\,\delta\phi\,,
\end{align}
where $\epsilon$ is the perturbation parameter representing the perturbation order of  background Minkowski tetrad. It is given by a four-dimensional identity matrix given by the Kronecker delta $\delta^{A}_{\mu}$. It is sufficient to expand the scalar field up to first order since higher order terms do not contribute. Additionally, it should be noted that the background scalar field can be taken to be as a function of time and apply the unitary gauge such that
\begin{align}
    \phi \rightarrow \phi(t)\,.
\end{align} 
Moreover, the perturbations of  arbitrary functions present in the Lagrangian are obtained by performing a Taylor expansion up to second order such that
\begin{align}
    &G_{i}(\phi,X) = G_{i} + \epsilon\,G_{i,X} X^{(1)} + \epsilon^{2}\left(\tfrac{1}{2}G_{i,XX} (X^{(1)})^{2} + G_{i,X} X^{(2)}\right)\,,\\
   &G_{\text{Tele}}(\phi,X,T,T_{\text{ax}},T_{\text{vec}},I_{2},J_{1},J_{3},J_{5},J_{6},J_{8},J_{10}) \nonumber \\
   & \qquad = G_{\text{Tele}} + \epsilon(G_{\text{Tele},X}X^{(1)} + G_{\text{Tele},I_{2}}I_{2}^{(1)}) + \tfrac{1}{2}\epsilon^{2}(G_{\text{Tele},XX}(X^{(1)})^{2}+ G_{\text{Tele},I_{2}I_{2}}(I_{2}^{(1)})^{2})\nonumber \\
   & \quad \qquad + \epsilon^{2}\Big(G_{\text{Tele},X} X^{(2)}  + G_{\text{Tele},T}T + G_{\text{Tele},T_{\text{ax}}} T_{\text{ax}} + G_{\text{Tele},T_{\text{vec}}} T_{\text{vec}} + G_{\text{Tele},I_{2}} I_{2}^{(2)} \Big) \,,
\end{align}
where, for $i=\{2,3,4,5\}$, $j=\{X,XX\}$ and  $k=\{X,T,T_{\text{ax}},T_{\text{vec}},XX,I_2I_2\}$, $G_{i,j}$ and $G_{\text{Tele},k}$ are background functions, such that $XX$ and $I_{2}I_{2}$ are the second order derivatives with respect to $X$ and $I_{2}$. The numbered superscripts represent the order of perturbation of the scalar invariant.

In this section, we consider a scalar-vector-tensor (SVT) decomposition of the tetrad based on the formalism applied in Ref.~\cite{Bahamonde:2020lsm} for first order perturbations:
\begin{align}\label{tetrad_perturbation}
    \delta\udt{e}{A}{\mu} \coloneqq \begin{bmatrix}
    \varphi &&&& -(\partial_{i}\beta + \beta_{i})\\
    \delta^{I}_{i}(\partial^{i}b + b^{i}) &&&& \delta^{Ii}\left(-\psi\delta_{ij} + \partial_{i}\partial_{j}h + \partial_{i}h_{j}+\partial_{j}h_{i} + \tfrac{1}{2}h_{ij} + \epsilon_{ijk}(\partial^{k}\sigma + \sigma^{k})\right)
    \end{bmatrix}\,,
\end{align}
where $\{\varphi,\beta,b,\psi,h\}$ are scalars and $\sigma$ is a pseudoscalar of 1 degree of freedom (DoF) each, $\{\beta_{i},b_{i},h_{i}\}$ are vectors and $\sigma_{i}$ is a pseudovector of 1 DoFs each, and $h_{ij}$ is the tensor mode of 2 DoFs, for a total of of 16 DoFs. The tensor modes are symmetric $h_{ij} = h_{(ij)}$, traceless $\delta^{ij}h_{ij} = 0$ and divergenceless $\partial^{i}h_{ij} = 0$. See also Ref. \cite{Abedi:2017jqx,Abedi:2018lkr}. The divergenceless property also applies for vectors and pseudovectors such that $\partial^{i}\alpha_{i}=0$ where $\alpha=\{\beta,b,h,\sigma\}$. Unlike Ref.~\cite{Wu:2012hs-no_antisymmetry,Chen:2010va-no_antisymmetry}, the pseudoscalar and pseudovector are included to account for the anti-symmetry of the tetrad. The mid-range Latin indices are spacial coordinates: $\{I,J,K,\ldots\}$ for spacial inner bundle and $\{i,j,k,\ldots\}$ for spacial spacetime manifold. Note, $\delta_{ij}$ is the spacial Minkowski metric such that $\delta_{ij}=-\eta_{ij}$. In turn, the first order metric perturbation from Eq.~\eqref{metric-definition} is given by
\begin{align}
    \delta g_{\mu\nu} = \begin{bmatrix}
    2\varphi &&&& \partial_{i}\mathcal{B} + \mathcal{B}_{i}\\
    \partial_{i}\mathcal{B} + \mathcal{B}_{i} &&&& 2\left(-\psi\delta_{ij} + \partial_{i}\partial_{j}h + \partial_{i}h_{j}+\partial_{j}h_{i} + \tfrac{1}{2}h_{ij}\right)
    \end{bmatrix}\,,
\end{align}
where $\mathcal{B} = -\beta+b$ and $\mathcal{B}_{i}=-\beta_{i}+b_{i}$. The off-diagonals are identical hence verifying the symmetry of the metric. The pseudoscalar and pseudovectors no longer play a role, and thus, the number of DoFs reduces to 10.

The gauge transformation through the  coordinate change \cite{Bruni:1996im,Malik:2008im,Nakamura:2006rk} is
\begin{align}
    \tilde{x}^{\mu} \rightarrow x^{\mu} + \xi^{\mu}\,,
\end{align}
where $\xi^{\mu}$ is a vector field  applied to study the gauge transformation of  perturbative quantities. Such a coordinate transformation can be extended to include orders higher than second one, but it is sufficient to consider up to this point. Thus, the transformations for first tetrad are given by~\cite{Hohmann:2020vcv}
\begin{align}
    \delta\udt{\tilde{e}}{A}{\mu} &= \delta\udt{e}{A}{\mu} + \mathcal{L}_{\xi_{(1)}} \udut{e}{A}{\mu}{(0)}\,,
\end{align}
where $\mathcal{L}_{\xi}$ is the Lie derivative along $\xi^{\mu}$ wherein $\xi^{\mu} = (\xi^{0},\xi^{i}+\delta^{ij}\partial_{j}\xi)$ further splits and once again obeys divergencelessness as $\partial_{i}\xi^{i}=0$. The analysis of each combination of temporal and spatial parts of the tetrad indicates that $\psi$, $\sigma$, $\beta_{i}$ and $h_{ij}$ are gauge invariant in Minkowski background while the rest have the following gauge transformations
\begin{align} \label{gauge_choice}
    \tilde{\varphi} &= \varphi - \dot{\xi}^{0}\,, \qquad \beta = \beta - \xi^{0}\,, \qquad \tilde{b} = b - \dot{\xi}\,, \qquad \tilde{h}=h-\xi\,, \nonumber \\
    \tilde{b}_{i} &= b_{i} + \dot{\xi}_{i}\,, \qquad \tilde{h}_{i} = h_{i} + \frac{1}{2}\xi_{i}\,, \qquad \tilde{\sigma}_{i} = \sigma_{i} - \frac{1}{2} \epsilon_{ijk} \partial^{j}\xi^{k}\,.
\end{align}
This shows that since the pseudoscalar $\sigma$ is gauge-invariant, it can be treated separately than the rest of the scalar modes, but the same cannot be said for the pseudovector. In fact, the pseudovector $\sigma_{i}$ can be expressed in terms of $h_{i}$ (and its derivative in terms of $b_{i}$). This implies that the vector and pseudovector cannot be decomposed~\cite{Izumi:2012qj}. 

Next, we construct groups of non-gauge invariant quantities: $\{\varphi,\beta\}$, $\{b, h\}$ and $\{b_{i},h_{i},\sigma_{i}\}$. For gauge choice, a quantity from each group is set to zero~\cite{Izumi:2012qj}. In particular, we will choose $\beta = 0$, $h=0$ and $\sigma_{i}=0$.

\section{Ghost and Laplacian Instabilities in Minkowski Background} \label{sec:instabilities}

Ghost instabilities stem from a negative kinetic term in the action associated to a propagating degree of freedom. A ghost mode can be determined by expanding the action up to second order perturbations about a background. Therefore, for a Lagrangian of the form $\mathcal{L} = \dot{\vec{\chi}}^{t} A \dot{\vec{\chi}} + \ldots$, we impose that the eigenvalues of $A$ should be positive to  eliminate  ghost modes. The procedure is applied in what follows. The second order perturbation of the action is obtained by separately applying the scalar, vector and tensor field perturbations given by Eq.~\eqref{tetrad_perturbation}. The dynamical fields in the action are identified, while a system of equations is obtained by varying the action with respect to the auxiliary fields. This system of equations is substituted back into the action to eliminate the non-dynamic fields~\cite{DeFelice:2010aj-f(R)_theories,Gonzalez-Espinoza:2021mwr}.

In order to obtain a gauge invariant action, the action is varied with respect to the temporal and spatial part of the vector field associated with the coordinate transformation, and imposing that this vanishes. Then, a diagonalized kinetic matrix is constructed and a constraint is generated for each entry. In general, these constraints could be time dependent: this feature arises from the background spacetime and the background quantities of fields. When looking at the propagating speeds of these modes, a positive definite value should be applied in order to ensure that the perturbation does not lead to an exponential growth~\cite{Kobayashi:2019hrl_HorndeskiReview} i.e. $c^{2}>0$, referred to as gradient or Laplacian instability. Hence, both ghost and gradient instabilities are checked for each mode in order to obtain the constraints which lead to a stable model. It is worth  noticing that, throughout the analysis, a high $k$ limit has been considered since ghost constraints are imposed in the high energy regime.  This fact leads to physical propagating speeds~\cite{Gumrukcuoglu:2016jbh,DeFelice:2009bx_Perfect-Fluid}.

In general, the following procedure will be followed. For action perturbed up to second order, the non-dynamical modes are identified. By varying  with respect to these modes and substituting back into the action yields an action of the form 
\begin{align}
    \mathcal{S}^{(2)} = \frac{1}{(2 \pi)^{3}} \int dt d^{3}k \left[ \dot{\chi} \textbf{K}(t,k) \dot{\chi}  - \chi (k^{2} \textbf{G}(t,k) + \textbf{M}(t,k))\chi - \chi \textbf{Q}(t,k) \dot{\chi} \right]\,,
\end{align}
where $k$ is the co-vector for spatial Fourier transformation, $\chi$ is the set of dynamical modes and $\textbf{K},\textbf{G},\textbf{M}$ and $\textbf{Q}$ are coefficient matrices. The ghost modes correspond to the entries of $\textbf{K}$, while  $\textbf{M}$ and $\textbf{K}$ are of order $\mathcal{O}(k)$ at high $k$ limit. The determinant
\begin{align}
    \text{det}(\omega^{2} \textbf{K}- i \omega \textbf{Q} - k^{2} \textbf{G}-\textbf{M}) = 0\,,
\end{align}
can be solved. Here $\omega^2 = k^{2} c^{2}$  is the temporal Fourier transformation   such that $c^{2}$ is the propagating speed of the mode from which the Laplacian condition can be determined.

\subsection{Tensor Perturbations}

The tensor mode perturbations, presented in Eq.~\eqref{tetrad_perturbation}, can be extended as
\begin{align}\label{tensor_pert_action}
    \udt{e}{A}{\mu} \rightarrow 
    \begin{bmatrix}
    1 & 0 \\
    0 & \delta_{ij} + \frac{1}{2} \epsilon h_{ij} + \frac{1}{8} \epsilon^2 h_{ik} \udt{h}{k}{j}
    \end{bmatrix}\,,
\end{align}
analogous to the Arnowitt-Deser-Misner (ADM) for  tensor modes~\cite{Gonzalez-Espinoza:2021mwr}. The action can be extended to second order perturbations such that
\begin{align}
    \mathcal{S}_{\text{T}}^{(2)} &= \frac{1}{2} \int d^{4}x \Big[ \mathcal{M}_{\text{T}}\, \dot{h}_{ij}\dot{h}^{ij} - \mathcal{N}_{\text{T}}\, \partial_{m}h_{ij}\partial^{m}h^{ij} + \mathcal{P} h_{ij}h^{ij}
   \Big]
\end{align}
where 
\begin{align}
    \label{M_tensor}\mathcal{M}_{\text{T}} &= G_{4} - 2 X G_{4,X} + X G_{5,\phi} - G_{\text{Tele},T} + \tfrac{1}{2} X G_{\text{Tele},J_{5}} + 2X G_{\text{Tele},J_{8}}\,,\\
    \label{N_tensor}\mathcal{N}_{\text{T}} &= G_{4} - X(G_{5,\phi} + \ddot{\phi} G_{5,X}) - G_{\text{Tele},T}\,,\\
    \label{P_tensor} \mathcal{P}_{\text{T}} &= -\tfrac{1}{2} \tfrac{d}{dt}(\dot{\phi}G_{\text{Tele},I_{2}})\,.
\end{align}
For the sake of simplicity, we switch to Fourier space such that
\begin{align}
    \mathcal{S}_{\text{T}}^{(2)} &= \tfrac{1}{2} \int dt \frac{d^{3}k}{(2\pi)^{3}} \Big[
    \mathcal{M}_{\text{T}} \dot{h}_{ij} \dot{h}^{ij} + \left(-k^{2}\mathcal{N}_{\text{T}} + \mathcal{P}_{\text{T}}\right)h_{ij}h^{ij}\,.\label{eq:tensor_action}
    \Big]
\end{align}
This result is obtained after applying integration by parts, removing the surface terms and applying the background Eqs.~(\ref{background-equations-1}-\ref{background-equations-3}). By imposing $\mathcal{M}_{\text{T}} > 0$, the theory is ghost free in  tensor modes. When considering a constant background scalar $\phi$, this implies that $G_{4}-G_{\text{Tele},T} > 0$ which corresponds to the condition $G_{4}-G_{\text{Tele},T} \neq 0$ imposed in Ref.~\cite{Bahamonde:2021dqn-GW} in order to ensure that there are tensor mode DoFs propagating. 
The propagation speed of tensor modes is given by
\begin{align}
    c_{\text{T}}^{2} = \frac{\mathcal{N}_{\text{T}}}{\mathcal{M}_{\text{T}}} = \frac{G_{4} - X(G_{5,\phi} + \ddot{\phi} G_{5,X}) - G_{\text{Tele},T}}{ G_{4} - 2 X G_{4,X} + X G_{5,\phi} - G_{\text{Tele},T} + \tfrac{1}{2} X G_{\text{Tele},J_{5}} + 2X G_{\text{Tele},J_{8}}}\,,\label{eq:tensor_total_speed}
\end{align}
for which $c_{\text{T}}^{2}> 0$ is required to ensure gradient stability. Hence, $\mathcal{M}_{\text{T}} > 0$ and $\mathcal{N}_{\text{T}} > 0$ result in ghost and gradient stability, respectively. The  same result would by obtained when eliminating $I_{2}$ contribution from the $G_{\text{Tele}}$ function. Following the observations of gravitational wave signal GW170817~\cite{LIGOScientific:2017vwq} and its electromagnetic counterpart GRB170817A~\cite{Goldstein:2017mmi}, it is interesting to calculate the deviation from speed of light propagation such that the excess speed is given by
\begin{align}
    \alpha_{\text{T}} = c^{2}_{\text{T}} - 1 = \frac{X(2G_{4,X}-2G_{5,\phi}-\ddot{\phi}G_{5,X}-\tfrac{1}{2}G_{\text{Tele},J_{5}}-2G_{\text{Tele},J_{8}})}{G_{4} - 2 X G_{4,X} + X G_{5,\phi} - G_{\text{Tele},T} + \tfrac{1}{2} X G_{\text{Tele},J_{5}} + 2X G_{\text{Tele},J_{8}}}\,,
    \label{eq:tensor_excess_speed}
\end{align}
which corresponds to the result obtained through the GW propagation equation~\cite{Bahamonde:2019ipm}, a value which  is highly constrained such that a graviton mass would be minute.

Additionally, from the general result of teleparallel  Horndeski analogue, given by Eq.~\eqref{tensor_pert_action}, one may obtain the results of well-studied theories from literature, as summarized in Table~\ref{tab:Tensor_Cases}. See also Ref.\cite{Capozziello:2018gms} for the metric cases derived from Noether symmetries.  As an extension analogous to standard Horndeski theory, the stability conditions can be obtained by setting $G_{\text{Tele}}=0$. The ghost modes are excluded for when the constant of the kinetic term is $G_{4}-2XG_{4,X}+X G_{5,X} >0$ while gradient stability is obtained for $G_{4}-X(G_{5,\phi}+\ddot{\phi}G_{5,X}) > 0$, in agreement with Refs.~\cite{Kobayashi:2019hrl_HorndeskiReview,DeFelice:2011bh-Horndeski-DEGalileon,Kobayashi:2011nu_Generalise-G-Inflation,Gao:2011qe_Non-Gaussianities} when taking the appropriate limits to Minkowski spacetime. An example of a subclass of Horndeski gravity is the Brans-Dicke theory, where $G_{2} = \tfrac{2 w_{\tiny{\text{BD}}} X}{\phi}$ and $G_{4}=\phi$ for which $w_{\tiny{\text{BD}}}$ is the Dicke coupling constant~\cite{Brans:1961sx}, other constants are set to vanish and ghost instabilities are avoided for $\phi > 0$. By considering the generalized Brans Dicke theory, ghost instabilities are avoided for a positive value of a function of the scalar field, $F(\phi)>0$~\cite{DeFelice:2010jn_Generalised-Brans-Dicke}. Another example is $f(\lc{R})$ where $G_{2}=f(\phi)-\phi f'(\phi)$, $G_{4}=f'(\phi)$, the rest of the constants are zero, implying that ghost stability is achieved for  $f'(\phi)>0$. A final subcase of Horndeski gravity considered here is  GR. In this case, the only non-vanishing constant is $G_{4}=1$ for which there are no ghost modes since $\mathcal{M}_{\text{T}} = 1 > 0$. All of these subcases do not result in any gradient instabilities as $c_{\text{T}} = 1 > 0$. 

Next, we look at cases that arise due to the inclusion of teleparallel terms. For a purely teleparallel theory such as $f(T)$ gravity, all terms are set to vanish except for $G_{\text{Tele},T} = f(T)$. This implies in $-f'(T)> 0$ to avoid ghost modes similar to the result obtained in Ref.~\cite{Izumi:2012qj}. An equivalent result is obtained for scalar-tensor theory with a Lagrangian of the form $\mathcal{L}=f(\phi,T)+X P(\phi)$, an extension of $f(T)$~\cite{Hohmann:2018rwf,Gonzalez-Espinoza:2021mwr}. Once again, gradient instabilities are not an issue. Finally, the case where only $I_{2}$ contributions are present is considered. In this case, all constants are going to vanish while $G_{\text{Tele},I_{2}}\neq 0$ leads to a non-dynamical degree of freedom.

\begin{table}[h!]
\footnotesize
    \centering
    \begin{tabular}{|c|c|c|c|}
    \hline
        \textbf{Theory} & \textbf{Case} & $\mathcal{M}_{\text{T}}$ & $\mathcal{N}_{\text{T}}$ \\
        \hline
        \multirow{2}{*}{Horndeski} & \multirow{2}{*}{$G_{\text{Tele}} = 0$} & \multirow{2}{*}{\makecell{$G_{4} - 2XG_{4,X} + \ddot{\phi} X G_{5,X}$}} & \multirow{2}{*}{\makecell{$G_{4} - X(G_{5,\phi}+\ddot{\phi}G_{5,X})$}}\\
        &&&\\
        \hline
        \multirow{2}{*}{\makecell{Generalized\\ Brans-Dicke}} & \multirow{2}{*}{\makecell{$G_{\text{Tele}}=G_{5}=0$, $G_{2}=B(\phi) X$\\ $G_{3}=2 \xi(\phi) X$, $G_{4}=\tfrac{1}{2}F(\phi)$}}  & \multirow{2}{*}{\makecell{$\tfrac{1}{2}F(\phi)$}} & \multirow{2}{*}{\makecell{$\tfrac{1}{2}F(\phi)$}}\\
        & & &\\
        \hline
        \multirow{2}{*}{Brans-Dicke} & \multirow{2}{*}{\makecell{$G_{\text{Tele}}=G_{3}=G_{5}=0$\\ $G_{2}=\tfrac{2w_{\tiny{\text{BD}}}X}{\phi}$, $G_{4}=\phi$}}  & \multirow{2}{*}{\makecell{$\phi$}} & \multirow{2}{*}{\makecell{$\phi$}}\\
        & & &\\
        \hline
        \multirow{2}{*}{$f(\lc{R})$} & \multirow{2}{*}{\makecell{$G_{\text{Tele}}=G_{3}=G_{5}=0$\\ $G_{2}=f(\phi)-\phi f'(\phi)$, $G_{4}=f'(\phi)$}} & \multirow{2}{*}{\makecell{$f'(\phi)$}} & \multirow{2}{*}{\makecell{$f'(\phi)$}}\\
        & & &\\
        \hline
        \multirow{2}{*}{General Relativity} & \multirow{2}{*}{\makecell{$G_{\text{Tele}}=G_2 = G_3 = G_5 = 0$\\ $G_{4}=1$ }} & \multirow{2}{*}{\makecell{$1$ }} & \multirow{2}{*}{$1$}\\
        & & &\\
        \hline
        \multirow{2}{*}{\makecell{Teleparallel\\ or $f(T)$}} & \multirow{2}{*}{\makecell{$G_{2}=G_{3}=G_{4}=G_{5}=0$,\\ $G_{\text{Tele}}=f(T)$}} & \multirow{2}{*}{\makecell{$-f'(T)$}} & \multirow{2}{*}{\makecell{ $-f'(T)$}}\\
        &&&\\
        \hline
        \multirow{2}{*}{$f(\phi,T) + X P(\phi)$}
        & \multirow{2}{*}{\makecell{$G_{3} = G_{4} = G_{5}=0$\\$G_{2}=XP(\phi)$, $G_{\text{Tele}}=f(\phi,T)$}}
        & \multirow{2}{*}{\makecell{$ -f_{,T}(\phi,T)$}} & \multirow{2}{*}{\makecell{$ -f_{,T}(\phi,T)$}}\\
        & & &\\
        \hline
        \multirow{2}{*}{$G_{\text{Tele},I_{2}}$ only} 
        & \multirow{2}{*}{\makecell{$G_{2}=G_3 = G_4 = G_5 = 0$\\ $G_{\text{Tele},T}=0$}} 
        & \multicolumn{2}{c|}{\multirow{2}{*}{\makecell{no propagating mode}}}\\
        &&\multicolumn{2}{c|}{}\\
        \hline
    \end{tabular}
    \caption{List of literature models with the respective ghost $\mathcal{M}_{\text{T}}$ and gradient stability $\mathcal{N}_{\text{T}}$ conditions are positive definite, while the propagation speed is $c_{\text{T}}=\mathcal{N}_{\text{T}}/\mathcal{M}_{\text{T}}$ for tensor modes. The models include Horndeski theory~\cite{Horndeski:1974wa,Kobayashi:2019hrl_HorndeskiReview,DeFelice:2011bh-Horndeski-DEGalileon,Kobayashi:2011nu_Generalise-G-Inflation}, Generalized Brans-Dicke~\cite{DeFelice:2010jn_Generalised-Brans-Dicke} and Brans-Dicke~\cite{Brans:1961sx}, $f(\mathring{R})$ theory, General Relativity, $f(T)$ theory~\cite{Izumi:2012qj}, $f(\phi,T)$ theory~\cite{Gonzalez-Espinoza:2021mwr} and the case where the action is dependent on $I_{2}$ only.}
    \label{tab:Tensor_Cases}
\end{table}

\subsection{Vector Perturbations}\label{sec:vector_pert}

Next, we consider the vector perturbation of the tetrad~\eqref{tetrad_perturbation} with the application of gauge fixing which eliminates the pseudovector and any coupling with it such that
\begin{align}
    \udt{e}{A}{\mu} \rightarrow 
    \begin{bmatrix}
        1 & -\epsilon\beta_{i}\\
        \epsilon\delta^{I}_{i} b^{i} & \delta^{Ii} (\delta_{ij} + \epsilon (\partial_{i}h_{j} + \partial_{j}h_{i})) 
    \end{bmatrix}\,.
\end{align}
This result is a case of only vector modes in this portion of  decomposition. Extending the action to second order vector perturbations, we obtain
\begin{align} \label{vector_action_V1}
    \mathcal{S}^{(2)}_{\text{V}} &= \int dt \frac{d^{3}k}{(2\pi)^{3}} \Big[  4k^{4}\mathcal{A} h_{i}h^{i} + \mathcal{E} \dot{\beta}_{i}\dot{\beta}^{i} \nonumber \\
    & \qquad \qquad \qquad  \qquad + k^{2}(4 \mathcal{B} \dot{h}_{i} (\dot{h}^{i} - b^{i}) + \mathcal{C} b_{i}b^{i} + \mathcal{D} \beta_{i}\beta^{i}  + 4 \mathcal{F} h_{i}\beta^{i} + 4 \mathcal{G} h_{i}\dot{\beta}^{i} - 2\mathcal{H} \beta_{i}b^{i}) \Big]\,,
\end{align}
where
\begin{align}
    \mathcal{A} &= G_{\text{Tele},T_{\text{vec}}} + \tfrac{1}{9} X \left(-G_{\text{Tele},J_{8}} + X G_{\text{Tele},J_{6}} - \tfrac{5}{2} G_{\text{Tele},J_{5}} - 3 G_{\text{Tele},J_3}\right)\,, \nonumber \\
    \mathcal{B} &= G_{4} - 2X G_{4,X} +X G_{5,\phi}- G_{\text{Tele},T} + X \left(2 G_{\text{Tele},J_{8}} + \tfrac{1}{2} G_{\text{Tele},J_{5}}\right)\,, \nonumber\\
    \mathcal{C} &= \mathcal{B} + \tfrac{1}{2} XG_{\text{Tele},J_{5}} + \tfrac{2}{3} X G_{\text{Tele},J_{10}} + \tfrac{2}{9}G_{\text{Tele},T_{\text{ax}}}\,, \nonumber\\
    \mathcal{D} &= \mathcal{C} + X G_{\text{Tele},J_{5}} - 2X G_{\text{Tele},J_{8}} - 2 X G_{\text{Tele},J_{10}}\,,\nonumber \\
    \mathcal{E} &= 4 \mathcal{A} - 3 G_{\text{Tele},T_{\text{vec}}} + 2 X G_{\text{Tele},J_{3}} \,, \nonumber \\
    \mathcal{F} &= \tfrac{1}{2}\dot{\phi} G_{\text{Tele},I_{2}} - \tfrac{d}{dt}\left(G_{4}-2XG_{4,X} + X G_{5,\phi}\right)\,.\nonumber\\
    \mathcal{G} &= 2\mathcal{A}-\mathcal{B} - 3 G_{\text{Tele},T_{\text{vec}}} + \tfrac{1}{3}X G_{\text{Tele},J_{3}} - 2 X G_{\text{Tele},J_{8}} + \tfrac{5}{2} XG_{\text{Tele},J_{5}}\,, \nonumber \\
    \mathcal{H} &= \mathcal{C} - 2 X G_{\text{Tele},J_{8}} - \tfrac{4}{9}G_{\text{Tele},T_{\text{ax}}}\,,
\end{align}
where $b_{i}$ is the only auxiliary vector mode. The variation with respect to this non-dynamical mode yields
\begin{align}
   0 &= \mathcal{H} \beta_{i} + 2 \mathcal{B}\dot{h}_{i} - \mathcal{C} b_{i}\,.
\end{align}
By solving for the auxiliary field and plugging it back  into Eq.~\eqref{vector_action_V1}, we get
\begin{align}\label{action_vector_V2}
    \mathcal{S}_{\text{V}}^{(2)} &= \int dt \frac{d^{3}k}{(2\pi)^{3}} \Bigg[
    \mathcal{E} \dot{\beta}_{i} \dot{\beta}^{i} + 4 k^{2} \frac{\mathcal{B}(\mathcal{C}-\mathcal{B})}{\mathcal{C}} \dot{h}_{i}\dot{h}^{i} -4 \frac{\mathcal{BH}}{\mathcal{C}} \beta_{i}\dot{h}^{i} + 4 \mathcal{G}h_{i}\dot{\beta}^{i} \nonumber \\
    & \qquad \qquad \qquad \qquad + 4 k^{4} \mathcal{A} h_{i}h^{i} + k^{2}\left( \frac{\mathcal{CD}-\mathcal{H}^{2}}{\mathcal{C}}\beta_{i}\beta^{i} + 4 \mathcal{F} \beta_{i}h^{i}\right)
    \Bigg]\,.
\end{align}
Applying integration by parts, the system  reduces to
\begin{align}\label{vector_action3}
     \mathcal{S}_{\text{V}}^{(2)} &= \int dt \frac{d^{3}k}{(2\pi)^{3}} \Bigg[
    \mathcal{E} \dot{\beta}_{i} \dot{\beta}^{i} + 4 k^{2} \frac{\mathcal{B}(\mathcal{C}-\mathcal{B})}{\mathcal{C}} \dot{h}_{i}\dot{h}^{i} + 4\left( \mathcal{G}+\frac{\mathcal{BH}}{\mathcal{C}}\right) h_{i}\dot{\beta}^{i} \nonumber \\
    & \qquad \qquad \qquad \qquad + 4 k^{4} \mathcal{A} h_{i}h^{i} + k^{2}\frac{\mathcal{CD}-\mathcal{H}^{2}}{\mathcal{C}}\beta_{i}\beta^{i} + 4k^{2}\left( \mathcal{F} + \frac{d}{dt}\left(\frac{\mathcal{BH}}{\mathcal{C}}\right)\right)\beta_{i}h^{i}
    \Bigg]\,.
\end{align}
Starting with the ghost stability conditions, we look at the kinetic terms in the action and impose
\begin{align}\label{M_vector}
    \mathcal{M}_{\text{V}}^{\beta} = \mathcal{E} > 0\,, \qquad \text{and} \qquad \mathcal{M}_{\text{V}}^{h} = 4k^{2}\frac{\mathcal{B}(\mathcal{C}-\mathcal{B})}{\mathcal{C}}>0.
\end{align}
Next, the action is fully transformed in Fourier space wherein $\partial_{0}\rightarrow -i\omega$ such that the Lagrangian can be written as 
\footnotesize\begin{align}
     \mathcal{L}_{\text{V}}^{(2)} &= 
     \begin{pmatrix}
     \beta_{i} & h_{i}
     \end{pmatrix}
     \begin{pmatrix}
         - \omega^{2} \mathcal{E} + k^{2} \frac{\mathcal{CD}-\mathcal{H}^{2}}{\mathcal{C}} & -2 i \omega \left(\mathcal{G}+\frac{\mathcal{BH}}{\mathcal{C}}\right) + 2 k^{2}\left(\mathcal{F}+\frac{d}{dt}\left(\frac{\mathcal{BH}}{\mathcal{C}}\right)\right)\\
         -2 i \omega \left(\mathcal{G}+\frac{\mathcal{BH}}{\mathcal{C}}\right) + 2 k^{2}\left(\mathcal{F}+\frac{d}{dt}\left(\frac{\mathcal{BH}}{\mathcal{C}}\right)\right) & -4 \omega^{2}k^{2} \frac{\mathcal{B}(\mathcal{C}-\mathcal{B})}{\mathcal{C}} + 4 k^{2} \mathcal{A}
     \end{pmatrix}
     \begin{pmatrix}
          \beta_{i}\\
          h_{i}
      \end{pmatrix}\,.
\end{align}
\normalsize By setting the determinant of the above matrix to zero and expanding to $\mathcal{O}(\tfrac{1}{k})$, the discriminant can be written as
\begin{align}
    \omega^{4} &+ \left(k^{2}\frac{\mathcal{A}\mathcal{C}^{2}\mathcal{E}+\mathcal{BC}(\mathcal{CD}-\mathcal{H}^{2}) + \mathcal{B}^{2}(-\mathcal{CD}+\mathcal{H}^{2})}{\mathcal{B}(\mathcal{C}-\mathcal{B})\mathcal{CE}} + \mathcal{O}(k^{0})\right)\omega^{2} \nonumber \\
     & \qquad \qquad \qquad \qquad \qquad \qquad \qquad + \left(k^{4}\frac{\mathcal{A}(-\mathcal{CD}+\mathcal{H}^{2})}{\mathcal{B}(\mathcal{B}-\mathcal{C})\mathcal{E}} + \mathcal{O}(k^{2})\right) = 0\,,
\end{align}
such that, by substituting the solution $\omega^{2} = k^{2} c_{\text{V}}$, yields 
\begin{align}
     \left(c_{\text{V}}^{2}\right)^{\beta} = \frac{\mathcal{H}^{2}-\mathcal{CD}}{\mathcal{CE}}>0\,, \qquad \qquad  \left(c_{\text{V}}^{2}\right)^{h} = -\frac{\mathcal{AC}}{\mathcal{B}(-\mathcal{B}+\mathcal{C})} >0\,.
\end{align}
It leads to the Laplacian conditions for the respective dynamical mode
\begin{align}~\label{N_vector}
    \mathcal{N}_{\text{V}}^{\beta} = \frac{\mathcal{H}^{2}-\mathcal{CD}}{\mathcal{C}}>0 \,, \qquad \qquad \mathcal{N}_{\text{V}}^{h}= -4 \mathcal{A}k^{2} >0\,.
\end{align}
These propagating modes are stemming out from the introduction of the $J_i$ terms in the BDLS action \eqref{teledeski-action}. They represent the quadratic contractions of the scalar and torsion fields $G_{\text{Tele},T_{\text{ax}}}$ and $G_{\text{Tele},T_{\text{vec}}}$.

\paragraph{Constant Background Scalar Field:} With regards to the scalar field, the unitary gauge is applied to the perturbative part of the scalar, while, at background level, it is a function of time only. Alternatively, if one considers a constant scalar field $\phi = c$, all terms in the actions are retained, hence the results of Eqs~\eqref{M_vector} and \eqref{N_vector} apply. It is important to notice that the  conditions contradict each other. This can be easily seen from the fact that $\mathcal{E} = \mathcal{A}$. For the no-ghost condition for  $\beta_{i}$, it is $\mathcal{A}>0$, while the no-Laplacian condition of $h_{i}$ gives $\mathcal{A}<0$. Moving forward, one might consider the case where $\mathcal{A}$ is eliminated from the action. In fact, none of the vector modes are dynamical, thus it corresponds  to the results in Ref.~\cite{Bahamonde:2021dqn} for the DoF and polarisation modes in the vector portion of the decomposition.

\paragraph{Vanishing Higher Order Derivatives:} In the case where higher order spatial and mixed derivatives are eliminated, it is $\mathcal{A}=\mathcal{B}=\mathcal{G}=0$. The Lagrangian becomes
\begin{align}
    \mathcal{L}_{\text{V}}^{\text{vanishing higher der}} = \mathcal{E}k^{2} \dot{\beta}_{i}\dot{\beta}^{i} + k^{2}\left( \mathcal{C} b_{i}b^{i} - 2 \mathcal{H} b_{i} \beta^{i} + 4 \mathcal{F} h_{i}\beta^{i} + \mathcal{D} \beta_{i}\beta^{i}\right)\,,
\end{align}
hence the variation with respect to $b_{i}$ and $h_{i}$ are respectively given by
\begin{align}
    0 &= \mathcal{C} b_{i} - \mathcal{H} \beta_{i}\,, \qquad \qquad
    0 = 4 k^{2} \mathcal{F} \beta_{i}\,.
\end{align}
For the latter condition, we will opt for the non-trivial solution $\mathcal{F}= 0$. Thus,
\begin{align}
    \mathcal{L}_{\text{V}}^{\text{vanishing higher der}} = \mathcal{E}k^{2} \dot{\beta}_{i}\dot{\beta}^{i} - k^{2} \left(\frac{\mathcal{H}^{2} - \mathcal{CD}}{\mathcal{C}}\right) \beta_{i}\beta^{i}\,,
\end{align}
shows that $\beta_{i}$  are the only propagating modes according to  the ghost and Laplacian conditions given by Eqs~\eqref{M_vector} and \eqref{N_scalar}. Since the action is undefined for $\mathcal{C}=0$, Laplacian instability is avoided for  $\mathcal{D} < 0$.

\paragraph[$G_{\text{Tele}}=0$]{\boldmath $G_{\text{Tele}}=0$}: When eliminating the teleparallel contribution, the remaining coefficients are all dependent on $\mathcal{B}$, such that the Lagrangian, after integrating by parts, becomes
\begin{align}
    \mathcal{L}_{\text{V}}^{G_{\text{Tele}}=0} = \mathcal{B} k^{2} (4 \dot{h}_{i}\dot{h}^{i} - 4 \dot{h}_{i}b^{i}+ b_{i}b^{i} + \beta_{i}\beta^{i} + 4 \dot{h}_{i}\beta^{i} - 2 \beta_{i} b^{i})\,,
\end{align}
where $b$ and $\beta_{i}$ are non-dynamical. In this case, the variation with respect to both auxiliary modes leads to the same solution
\begin{align}
    0 = \mathcal{B}\left(b_{i} - \beta_{i} - 2 \dot{h}_{i}\right)\,.
\end{align}
For both solutions $\mathcal{B} = 0$ and $b_{i} = \beta_{i} + 2 \dot{h}_{i}$, it results in a zero Lagrangian.

\paragraph[$\mathcal{C} = 0$]{\boldmath $\mathcal{C} = 0$}: Eq.~\eqref{vector_action3} is undefined for $\mathcal{C} = 0$. Hence, substituting this condition into Eq.~\eqref{vector_action_V1} and varying with respect to the auxiliary field $b$, it  results in
\begin{align}
    0 &= 2 \mathcal{B} \dot{h}_{i} + \mathcal{H} \beta_{i}\,.
\end{align}
Solving for $\dot{h}_{i}$, the action has only $\beta_{i}$ as a dynamical mode, hence, varying with respect to $h_{i}$, yields
\begin{align}
    0 &=2 k^{2} \mathcal{A} h_{i} + \mathcal{F} \beta_{i} + \mathcal{G} \dot{\beta}_{i}\,,
\end{align}
such that
\begin{align}
    \mathcal{L}_{\text{V}}^{\mathcal{C}=0} = \frac{\mathcal{AE}-\mathcal{G}^{2}}{\mathcal{A}} \dot{\beta}_{i}\dot{\beta}^{i} + k^{2}\frac{\mathcal{BD}+\mathcal{H}^{2}}{\mathcal{B}} \beta_{i}\beta^{i} + \left[ -\frac{\mathcal{F}^{2}}{\mathcal{A}} + \frac{d}{dt} \left( 
\frac{\mathcal{FG}}{\mathcal{A}} \right)\right] \beta_{i} \beta^{i}\,.
\end{align}
By following the same procedure as done in the general case, ghost instability is avoided for
\begin{align}
    \mathcal{M}_{\text{V}}^{\mathcal{C}=0} = \frac{\mathcal{AE} - \mathcal{G}^{2}}{\mathcal{A}} > 0\,,
\end{align}
and the Laplacian stability condition, provided by the leading order in $k$, is given by
\begin{align}
    -\frac{\mathcal{BD}+\mathcal{H}^{2}}{\mathcal{B}} >0\,,
\end{align}
wherein the cases for vanishing $\mathcal{A}$ and $\mathcal{B}$ are discussed for vanishing higher order derivatives.

\paragraph[Vanishing $J_i$ terms:]{Vanishing \boldmath $J_i$ and $G_{\text{Tele},T_{\text{ax}}}$ terms:} In the case  $J_{1} = J_{3} = J_{5} = J_{6} = J_{10} = G_{\text{Tele},T_{\text{ax}}} = 0$, the Lagrangian results in the form 
\begin{align}
    \mathcal{L}_{\text{V}}^{J_{i} = G_{\text{Tele},T_{\text{ax}}}=0} = -\frac{\mathcal{B}(\mathcal{A}-\mathcal{B})}{\mathcal{A}} \dot{\beta}_{i} \dot{\beta}^{i} + \mathcal{O}(k^{(0)})\,.
\end{align}
Here there is a no-ghost condition, then for the propagating mode $\beta_{i}$, the propagating speed cannot be determined.

\paragraph[Vanishing $J_i$ terms:]{Vanishing \boldmath $J_i$ and $G_{\text{Tele},T_{\text{Vec}}}$ terms:} In the case  $J_{1} = J_{3} = J_{5} = J_{6} = J_{10} = G_{\text{Tele},T_{\text{vec}}}=0$, the action renders the $h_{i}$ modes  non-dynamical upon substituting the set of equations arising from the non-dynamical modes. The Lagrangian is written as
\begin{align}
     \mathcal{L}_{\text{V}}^{J_{i} = G_{\text{Tele},T_{\text{vec}}}=0} = \frac{4 k^{2}\mathcal{B}(2\mathcal{B}-\mathcal{C}+\mathcal{H})}{\mathcal{H}-\mathcal{C}} \dot{h}_{i} \dot{h}^{i} - k^{2}\left[ \frac{4 \mathcal{C} (\mathcal{F}+\dot{\mathcal{B}})^{2}}{\mathcal{C}^{2}-\mathcal{H}^{2}} + 4 \frac{d}{dt}\left(\frac{\mathcal{B}(\mathcal{F}+\dot{\mathcal{B}})}{\mathcal{H}-\mathcal{C}}\right) \right]h_{i}h^{i}\,.
\end{align}
the ghost condition, in terms of $k$, is given by
\begin{align}
    \mathcal{M}_{\text{V}}^{J_{i} = G_{\text{Tele},T_{\text{vec}}} = 0} = \frac{4k^{2}\mathcal{B}(2\mathcal{B} - \mathcal{C}+\mathcal{H})}{\mathcal{H}-\mathcal{C}} >0\,,
\end{align}
but the propagation speed is of order $\mathcal{O}(\frac{1}{k^{2}})$ which, in the high $k$ limit, is equivalent to zero. In fact, this can be further verified by varying with respect to the dynamical mode $h_{i}$ since it is not coupled with other modes. It is
\begin{align}
    \ddot{\psi} - \left( \frac{\frac{d}{dt}\left(\frac{\mathcal{B}(\mathcal{F}+\dot{\mathcal{B}})}{\mathcal{H}-\mathcal{C}}\right) \left(\mathcal{C}^{2}-\mathcal{H}^{2}\right) + \mathcal{C}\left(\mathcal{F}+\dot{\mathcal{B}}\right)^{2}}{\mathcal{B}(2\mathcal{B} - \mathcal{C}+\mathcal{H})(\mathcal{C}+\mathcal{H})} \right) \psi = 0\,,
\end{align}
where there is no coefficient for $k^{2}\psi$.

\subsection{Scalar Perturbations}

Let us now consider  the scalar modes. Since there is no mixing between the scalar and pseudoscalar modes, we will treat them separately. The tetrad perturbation using scalar modes is given by
\begin{align}
    \udt{e}{A}{\mu} \rightarrow
    \begin{bmatrix}
    1+\varphi & 0\\
    \delta^{I}_{i} \partial^{i}b & (1- \psi) \delta^{Ii} \delta_{ij}\,.
    \end{bmatrix}
\end{align}
As already stated, for the gauge invariant quantities, given by Eq.~\eqref{gauge_choice}, the gauge choice is  $\beta = h = 0$. The pseudoscalar will be analyzed separately. It should be noted that this situation is  identical to the Arnowitt-Deser-Misner (ADM) decomposition used in torsion-based theories~\cite{Gonzalez-Espinoza:2021mwr}. It leads to the same metric as that applied to curvature-based theories~\cite{Kobayashi:2019hrl_HorndeskiReview}. On the other hand, the choice  $\beta = b = 0$  results in the Newtonian (longitudinal) gauge. Substituting the tetrad perturbation to the action, followed by integration by parts and application of the background equations~(\ref{background-equations-1}-\ref{background-equations-3}), the second order perturbation of the action is given by
\begin{align} \label{scalar_action_1}
    \mathcal{S}_{\text{S}}^{(2)} &= \int dt \frac{d^{3}k}{(2\pi)^{3}} \Big[
    \mathcal{\bar{A}} \varphi^{2} + 6 \mathcal{\bar{D}} \psi^{2} - 6 \mathcal{\bar{F}} \dot{\psi}^{2} - 6 \mathcal{\bar{G}} \varphi \dot{\psi} \nonumber \\
    & \qquad \qquad \qquad \qquad - k^{2}(-\mathcal{\bar{B}} \varphi^{2} - 2 \mathcal{\bar{E}} \psi^{2} + 4\mathcal{\bar{H}} \varphi \psi - 2 \mathcal{\bar{G}} \varphi b - 4 \mathcal{\bar{F}} \dot{\psi}b) + k^{4}\mathcal{\bar{C}} b^{2}
    \Big]
\end{align}
where
\begin{align}
    \mathcal{\bar{A}} &=  X G_{2,X} + 2X^{2}G_{2,XX}-2XG_{3,\phi} - 2X^{2}G_{3,\phi X} + X G_{\text{Tele},X} + 2X^{2}G_{\text{Tele},XX}\,,\nonumber\\
    \mathcal{\bar{B}} &= G_{\text{Tele},T_{\text{vec}}} + \tfrac{2}{9}X\left(2 G_{\text{Tele},J_{8}} + 2 X G_{\text{Tele},J_{6}} - 5 G_{\text{Tele},J_{5}} + 3G_{\text{Tele},J_{3}}\right)\,,\nonumber\\
    \mathcal{\bar{C}} &= -G_{\text{Tele},T_{\text{vec}}} + X G_{\text{Tele},I_{2}I_{2}} + \tfrac{1}{3}X(4G_{\text{Tele},J_{8}}+G_{\text{Tele},J_{5}})\,,\nonumber\\
    \mathcal{\bar{D}} &= \tfrac{d}{dt}(\dot{\phi} G_{\text{Tele},I_{2}})\,,\nonumber\\
    \mathcal{\bar{E}} &= G_{4}-X (G_{5,\phi} - \ddot{\phi} G_{5,X}) -G_{\text{Tele},T} + 2 G_{\text{Tele},T_{\text{vec}}}\nonumber\\
    &+ \tfrac{1}{9}X\left(-2 G_{\text{Tele},J_{8}} + 2 X G_{\text{Tele},J_{6}} - 5G_{\text{Tele},J_{5}} - 6G_{\text{Tele},J_{3}}\right)\,,\nonumber\\
    \mathcal{\bar{F}} &= G_{4}-2X G_{4,X}+XG_{5,\phi} - G_{\text{Tele},T} + \tfrac{3}{2}(G_{\text{Tele},T_{\text{vec}}} - X G_{\text{Tele},I_{2}I_{2}})\,,\nonumber\\
    \mathcal{\bar{G}} &= -\dot{\phi} X G_{3,X} + \dot{\phi}G_{4,\phi} + 2 \dot{\phi} X G_{4,\phi X} + \tfrac{1}{2} \dot{\phi} G_{\text{Tele},I_{2}} + \dot{\phi} X G_{\text{Tele},XI_{2}}\,, \nonumber\\
    \mathcal{\bar{H}} &= G_{4}-2X G_{4,X}+XG_{5,\phi} - G_{\text{Tele},T}  + G_{\text{Tele},T_{\text{vec}}} \nonumber\\
    &+ \tfrac{1}{9}X(2G_{\text{Tele},J_{8}} - 2X G_{\text{Tele},J_{6}} + 5G_{\text{Tele},J_{5}}+\tfrac{3}{2}G_{\text{Tele},J_{3}}) \,.
\end{align}
Here $k$ is the co-vector. It is clear that variables $\varphi$ and $b$ are non-dynamical with respect to the time component. By varying with respect to each of these auxiliary fields, one obtains a set of equations
\begin{align}\label{eq:variation_scalar1}
    0 &= k^{2}(\bar{\mathcal{G}} b + \bar{\mathcal{B}}\varphi-2 \bar{\mathcal{H}}\psi) + \bar{\mathcal{A}}\varphi-3\bar{\mathcal{G}}\dot{\psi}\,,\\
    0 &=k^{2}\bar{\mathcal{C}}b + \bar{\mathcal{G}}\varphi + 2\bar{\mathcal{F}}\dot{\psi}\,,
\end{align}
By solving these equations of motion for the modes $\varphi$ and $b$, it leads to an action expressed in terms of the dynamical mode only. Considering the higher $k$ limit~\cite{Gonzalez-Espinoza:2021mwr}, that is
\begin{align} \label{eq:scalar_pert_action}
    \mathcal{S}_{\text{S}}^{(2)} &= \int dt \frac{d^{3}k}{(2\pi)^{3}} \Big[ -\frac{2\bar{\mathcal{F}}(3\bar{\mathcal{C}}+2\bar{\mathcal{F}})}{\bar{\mathcal{C}}} \dot{\psi}^{2} + k^{2} \frac{2(\bar{\mathcal{B}}\bar{\mathcal{E}}-2 \bar{\mathcal{H}}^{2})}{\mathcal{B}} \psi^{2} \nonumber \\
    & \qquad \qquad \qquad +2 \left( \frac{3\bar{\mathcal{B}}^{2}\bar{\mathcal{C}}\bar{\mathcal{D}}+2(\bar{\mathcal{A}}\bar{\mathcal{C}}-\bar{\mathcal{G}}^{2})\bar{\mathcal{H}}^{2}}{\bar{\mathcal{B}}^{2}\bar{\mathcal{C}}} + \frac{d}{dt}\left(\frac{\bar{\mathcal{B}}(3\bar{\mathcal{C}}+2\bar{\mathcal{F}})\bar{\mathcal{G}}\bar{\mathcal{H}}}{\bar{\mathcal{B}}^{2}\bar{\mathcal{C}}^{2}}\right)\right)\psi^{2}\Big]\,,
\end{align}
it results in an action depending on  the gauge invariant $\psi$. In the case of teleparallel analogue of Horndeski theory, the ghost stability is obtained for
\begin{align}
    \label{M_scalar}\mathcal{M}_{\text{S}} &= -\frac{2\bar{\mathcal{F}}(3\bar{\mathcal{C}}+2\bar{\mathcal{F}})}{\bar{\mathcal{C}}} > 0\,.
\end{align}
For the propagating speed, the temporal derivatives in the action are changed to the Fourier space i.e. $\partial_{0}\rightarrow -i \omega$ such that the dispersion relation is given by $\omega^{2} = k^{2} c_{\text{S}}^{2}$ for the scalar propagating speed $c_{\text{S}}^{2}$ since there are no higher order spatial derivatives in the action. By setting the determinant of the Lagrangian to zero such as
\begin{align}
    -\omega^{2} K+M = 0\,,
\end{align}
where $K$ is coefficient of $\dot{\psi}^{2}$ and $M$ is coefficient of $\psi^{2}$, the propagating speed at high $k$ limit is given by
\begin{align}\label{eq:scalar_DoF_speed}
    c_{\text{S}}^{2} &= \frac{-2\bar{\mathcal{C}}+\bar{\mathcal{B}}\bar{\mathcal{C}}\bar{\mathcal{E}}}{\bar{\mathcal{B}}\bar{\mathcal{F}}(3\bar{\mathcal{C}}+2\bar{\mathcal{F}})} >0\,,
\end{align}
where, by definition of $c_{\text{S}}^{2} = \frac{\mathcal{N}_{\text{S}}}{\mathcal{M}_{\text{S}}}$, the Laplacian stability is obtained for
\begin{align}\label{N_scalar}
  2 \frac{-\bar{\mathcal{B}}\bar{\mathcal{E}} + 2 \bar{\mathcal{H}}^{2}}{\bar{\mathcal{B}}} > 0\,,
\end{align}
which, within the high $k$ limit, is equivalent to consider the part of the action with coefficient of $k^{2}\psi^{2}$. It should be noted that the action in Eq.~\eqref{eq:scalar_pert_action} is particularly undefined for the case where $\bar{\mathcal{B}}$ and $\bar{\mathcal{C}}$ vanish, the latter expression resulting from considering higher order spatial derivatives in the Lagrangian. Hence, it is  interesting to carryout the analysis for such cases. By setting such expression to vanish in Eq.~\eqref{scalar_action_1} the action reduces to
\begin{align} \label{scalar_action_2}
    \mathcal{S}_{\text{S}}^{(2)} &= \int dt \frac{d^{3}k}{(2\pi)^{3}} \Big[
    \mathcal{\bar{A}} \varphi^{2} + 6 \mathcal{\bar{D}} \psi^{2} - 6 \mathcal{\bar{F}} \dot{\psi}^{2} - 6 \mathcal{\bar{G}} \varphi \dot{\psi} - k^{2}(- 2 \mathcal{\bar{E}} \psi^{2} + 4\mathcal{\bar{H}} \varphi \psi - 2 \mathcal{\bar{G}} \varphi b - 4 \mathcal{\bar{F}} \dot{\psi}b)
    \Big]\,.
\end{align}
Variations with respect to the auxiliary fields are slightly altered to
\begin{align}
    0 &= \bar{\mathcal{A}} \varphi - 3 \bar{\mathcal{G}} \dot{\psi} - 2k^{2}(2 \bar{\mathcal{H}} \psi - \bar{\mathcal{G}}b)\,,\\
    0 &= \bar{\mathcal{G}}\varphi + 2 \bar{\mathcal{F}} \dot{\psi}\,,
\end{align}
yielding the action upon substitution as
\begin{align}
    \mathcal{S}_{\text{S}}^{(2)} &= \int dt \frac{d^{3}k}{(2\pi)^{3}} \Big[
   + \left(\frac{4\mathcal{\bar{A}}\mathcal{\bar{F}}^{2}}{\mathcal{\bar{G}}^{2}}+6\mathcal{\bar{F}}\right)\dot{\psi}^{2} - k^{2}\left(-2\mathcal{\bar{E}}+4 \frac{d}{dt}\left(\frac{\mathcal{\bar{F}\bar{H}}}{\mathcal{\bar{G}}}\right)\right)\psi^{2} + 6\mathcal{\bar{D}}\psi^{2}
    \Big]\,.
\end{align}
to which the ghost and Laplacian conditions become
\begin{align} \label{M_scalar2}
    \mathcal{M}_{\text{S}} &= \frac{4\mathcal{\bar{A}}\mathcal{\bar{F}}^{2}}{\mathcal{\bar{G}}^{2}} 
 + 6 \mathcal{\bar{F}} >0\,,\\
    \label{N_scalar2}\mathcal{\bar{N}}_{\text{S}} &= -2\mathcal{\bar{F}}+4\frac{d}{dt}\left(\frac{\mathcal{\bar{F}\bar{H}}}{\mathcal{\bar{G}}}\right) >0\,.
\end{align}
 Different subcases are reported in Table~\ref{tab:Scalar_Cases}. Through the substitution of Eqs~\eqref{M_scalar} and \eqref{N_scalar},  conditions to avoid ghost and Laplacian instabilities are realised. Unless the considered Lagrangians  contain non-vanishing $\bar{\mathcal{B}}$ and $\bar{\mathcal{C}}$, which are terms arising from the teleparallel portion of the BDLS action, Eq.~\eqref{M_scalar2} and \eqref{N_scalar2} can be used.

\paragraph{Horndeski:}
The standard Horndeski case has identical expressions for  ghost and gradient stability conditions as those for teleparallel Horndeski with $\bar{\mathcal{B}}=\bar{\mathcal{C}}=\bar{\mathcal{D}}=0$ ~(\ref{M_scalar2}-\ref{N_scalar2}) with the difference that each constant does not have teleparallel contributions, implying and $\bar{\mathcal{H}}=\bar{\mathcal{F}}$.

\paragraph{Generalized Brans-Dicke:} Considering Generalised Brans-Dicke (GBD) theory means that $G_{2}=B(\phi)X$, $G_{3}=2X\xi(\phi)$, $G_{4}=\tfrac{1}{2}F(\phi)$ and all other terms are set to zero. When we take the background equations
\begin{align}
    B - 4 X \xi'=0\,, \qquad \text{and} \qquad
    \dot{\phi}^{2}F'' + \ddot{\phi}(F' - 2 \dot{\phi}^{2}\xi) = 0\,,
\end{align}
for ghost stability, we have
\begin{align}
    \mathcal{M}_{\text{S}}^{\text{GBD}} 
    &= \frac{ F [3 F'^{2}- 4\dot{\phi}^{2}(F\xi' + 3\xi F') + 12 \dot{\phi}^{4}\xi^{2} ]}{[ F'-2\dot{\phi}^{2}\xi]^{2}} > 0\,,
\end{align}
where $'$ denotes a derivative with respect to the respective variable of the function. In this case, it is $\phi$, while the gradient condition is
\begin{align}
    \mathcal{N}_{\text{S}}^{\text{GBD}} 
    &= \frac{F[3 F'^{3} - 2\dot{\phi}^{2}(5F'^{2}\xi - 4\xi F F'' - 2 F \xi' F') + 4\dot{\phi}^{4}(F'\xi^{2}-2F\xi'\xi) + 8\dot{\phi}^{6} \xi^{3}]}{[F'-2\dot{\phi}^{2}\xi]^{3}} > 0\,.
\end{align}
With these conditions, the speed of propagation is positive. It is
\begin{align}
(c_{\text{S}}^{\text{GBD}})^{2} &= \frac{3 F'^{3} - 2\dot{\phi}^{2}(5F'^{2}\xi - 4\xi F F'' - 2 F \xi' F') + 4\dot{\phi}^{4}(F'\xi^{2}-2F\xi'\xi) + 8\dot{\phi}^{6} \xi^{3}}{ [3 F'^{2} - 4\dot{\phi}^{2}(F\xi' + 3\xi F') + 12 \dot{\phi}^{4}\xi^{2}][F'-2\dot{\phi}^{2}\xi]}\,,
\end{align}
which correlates with  results obtained in Ref.~\cite{DeFelice:2010jn_Generalised-Brans-Dicke} for a Minkowski background. 

\paragraph{Brans-Dicke:} As for the Brans-Dicke theory  in Ref.~\cite{Brans:1961sx} with $G_{2}=\tfrac{2w_{\text{\tiny{BD}}}X}{\phi}$ and $G_{4}=\phi$, ghosts instabilities are avoided if $2\phi(3+2w_{\text{\tiny{BD}}})>0$. Provided that $\phi>0$, this implies that the Brans-Dicke constant is $w_{\text{\tiny{BD}}}>-\tfrac{3}{2}$~\cite{DeFelice:2010aj-f(R)_theories}. The gradient stability condition is given by $2 \phi(3 - 2 \phi \ddot{\phi}/\dot{\phi}^{2})>0$ to ensure a positive propagating speed. Moreover, when applying the second Friedman Eq.~\eqref{background-equations-2}, gradient condition reduces to $2\phi(3+w_{\text{\tiny{BD}}})$ such that $w_{\text{BD}}>-3$ and a unitary propagating speed~\cite{Ozer:2021qjb_Brans-Dicke-GW}, analogous to  $\xi = 0$, reported in Ref.~\cite{DeFelice:2010jn_Generalised-Brans-Dicke}, is obtained. The first Friedman Eq.~\eqref{background-equations-1} shows that the only non-trivial solution is that for $w_{\text{BD}}=0$, which results in identical ghost and gradient expressions.  

\paragraph[$f(\mathring{R})$:]{\boldmath $f(\mathring{R})$:} The scalar-tensor theory equivalent to $f(R)$~\cite{Teyssandier:1983zz-f(R)_scalar_tensor} is given by setting $G_{2} = f(\phi)-\phi f'(\phi)$ and $G_{4}=f'(\phi)$ while the rest of the functions are set  to vanish. See also Ref. \cite{Capozziello:2018gms}. In order to have ghost stability, it has to be $6 f'(\phi) > 0$, i.e. $f'(\phi)>0$. Once again, upon applying the background Eqs.~(\ref{background-equations-1}-\ref{background-equations-2}), the gradient condition reduces to $f'(\phi)>0$. In fact, as a subcase of GBD theory, the case  $\xi=0$  always results in a unitary speed propagating mode.

\paragraph[GR, $f(T)$, $f(\phi,T)$:]{\boldmath GR, $f(T)$, $f(\phi,T)$:} When considering the cases of GR, $f(T)$ and $f(\phi,T)$, the substitution in Eqs\,\~(\ref{M_scalar2}-\ref{N_scalar2}) results in an undefined expression. In these cases, although taking the appropriate limits of each coefficient does lead to a positive ghost condition value, the propagating speed is negative, thus scalar modes are not viable. This is expected in GR. Additionally, Ref.~\cite{Izumi:2012qj} shows that the only propagating scalar field is that dependent on the perturbation of  scalar field $\phi$, while, in this paper, we are applying the unitary gauge. As in Ref.~\cite{Gonzalez-Espinoza:2021mwr}, $f(\phi,T)$ theory gives rise to a propagating mode when considering cosmological perturbations. The same applies for GR with an additional canonical scalar field such that $G_{2}=X-V(\phi)$, $G_{4}=M_{\text{Pl}}^{2}/2$, $G_{3}=G_{5}=G_{\text{Tele}}=0$~\cite{Kobayashi:2019hrl_HorndeskiReview}.

\paragraph{Teleparallel:} If one considers only the teleparallel portions of the action, all terms from Eq.~\eqref{scalar_action_1} are retained, with all standard Horndeski contributions set to vanish. This implies that the ghost and Laplacian instabilities coincide with the form of Eqs.~\eqref{M_scalar} and~\eqref{N_scalar}, respectively, provided that the background equations are satisfied.

\begin{sidewaystable}[]
\footnotesize
    \centering
    \begin{tabular}{|c|c|c|c|}
    \hline
        \textbf{Theory} & \textbf{Case} & $\mathcal{M}_{\text{S}}$ & $\mathcal{N}_{\text{S}}$ \\
        \hline
        \multirow{2}{*}{Horndeski} 
        & \multirow{2}{*}{$G_{\text{Tele}} = 0$} 
        & \multirow{2}{*}{\makecell{$ 6 \mathcal{\bar{F}} + \frac{4\mathcal{\bar{A}}\mathcal{\bar{F}}^{2}}{\mathcal{\bar{G}}^{2}}$}} 
        &\multirow{2}{*}{\makecell{$ -2\mathcal{\bar{E}} + 4 \frac{d}{dt}(\frac{\mathcal{\bar{F}}^{2}}{\mathcal{\bar{G}}})$}} \\
        & & &\\ 
        \hline
        \multirow{3}{*}{\makecell{Generalized \\Brans-Dicke}} 
        & \multirow{3}{*}{\makecell{$G_{\text{Tele}}=G_{5}=0$,\\$G_{2}=B(\phi) X$, $G_{3}=2 \xi(\phi) X$,\\ $G_{4}=\tfrac{1}{2}F(\phi)$}} 
        & \multirow{3}{*}{\makecell{$\frac{ F [3 F'^{2}- 4\dot{\phi}^{2}(F\xi' + 3\xi F') + 12 \dot{\phi}^{4}\xi^{2} ]}{[ F'-2\dot{\phi}^{2}\xi]^{2}}$ }} 
        & \multirow{3}{*}{\makecell{$\frac{F[3 F'^{3} - 2\dot{\phi}^{2}(5F'^{2}\xi - 4\xi F F'' - 2 F \xi' F') + 4\dot{\phi}^{4}(F'\xi^{2}-2F\xi'\xi) + 8\dot{\phi}^{6} \xi^{3}]}{[F'-2\dot{\phi}^{2}\xi]^{3}}$ }} \\
        & & &\\
        & & &\\
        \hline
        \multirow{2}{*}{Brans-Dicke} 
        & \multirow{2}{*}{\makecell{$G_{\text{Tele}}=G_{3}=G_{5}=0$\\ $G_{2}=\tfrac{2w_{\text{\tiny{BD}}}X}{\phi}$, $G_{4}=\phi$}}  
        & \multirow{2}{*}{\makecell{$2\phi(3+2w_{\text{\tiny{BD}}})$}} 
        & \multirow{2}{*}{\makecell{$2 (3\phi - \frac{\ddot{\phi}}{X}) \rightarrow 2\phi(3+2w_{\text{\tiny{BD}}})$ }} \\
        & & &\\
        \hline
        \multirow{3}{*}{$f(\lc{R})$} 
        &\multirow{3}{*}{\makecell{$G_{\text{Tele}}=G_{3}=G_{5}=0$ \\ $G_{2}=f(\phi)-\phi f'(\phi)$,\\ $G_{4}=f'(\phi)$}} & \multirow{3}{*}{\makecell{$ 6 f'(\phi)$}} & \multirow{3}{*}{\makecell{$ -2f'(\phi) + 4 \frac{d}{dt}(\frac{f'(\phi)^{2}}{\dot{\phi} f''(\phi)}) \rightarrow 6 f'(\phi)$}}\\
        &&&\\
        &&&\\
        \hline
        \multirow{2}{*}{\makecell{GR}} & \multirow{2}{*}{\makecell{$G_{\text{Tele}}=G_2 = G_3 = G_5 = 0$\\ $G_{4}=1$}} & \multicolumn{2}{c|}{\multirow{2}{*}{\makecell{no propagating mode}}}\\
        &&\multicolumn{2}{c|}{}\\
        \hline
        \multirow{2}{*}{$f(T)$} 
        & \multirow{2}{*}{\makecell{$G_{2}=G_{3}=G_{4}=G_{5}=0$,\\ $G_{\text{Tele}}=f(T)$}} & \multicolumn{2}{c|}{\multirow{2}{*}{\makecell{no propagating mode}}}\\
        &&\multicolumn{2}{c|}{}\\
        \hline
        \multirow{3}{*}{$f(\phi,T) + X P(\phi)$}
        & \multirow{3}{*}{\makecell{$G_{3} = G_{4} = G_{5}=0$,\\$G_{2}=XP(\phi)$,\\ $G_{\text{Tele}}=f(\phi,T)$ }} & \multicolumn{2}{c|}{\multirow{3}{*}{\makecell{no propagating mode}}}\\
        &&\multicolumn{2}{c|}{}\\
         &&\multicolumn{2}{c|}{}\\
        \hline
        \multirow{2}{*}{Teleparallel} 
        & \multirow{2}{*}{\makecell{$G_{2}=G_3 = G_4 = G_5 = 0$}} 
        & \multirow{2}{*}{$-\frac{2\bar{\mathcal{F}}(3\bar{\mathcal{C}}+2\bar{\mathcal{F}})}{\bar{\mathcal{C}}}$} & \multirow{2}{*}{$- \frac{2(\bar{\mathcal{B}}\bar{\mathcal{E}}-2 \bar{\mathcal{H}}^{2})}{\mathcal{B}}$}\\
        &&&\\
        \hline
    \end{tabular}
    \caption{List of literature models with the respective ghost $\mathcal{M}_{\text{S}}$ and gradient stability $\mathcal{N}_{\text{S}}$ conditions are positive definite, and propagation speed $c_{\text{S}}=\mathcal{N}_{\text{S}}/\mathcal{M}_{\text{S}}$ for scalar modes. The models include Horndeski theory~\cite{Horndeski:1974wa,Kobayashi:2019hrl_HorndeskiReview,DeFelice:2011bh-Horndeski-DEGalileon,Kobayashi:2011nu_Generalise-G-Inflation}, Generalized Brans-Dicke~\cite{DeFelice:2010jn_Generalised-Brans-Dicke} and Brans-Dicke~\cite{Brans:1961sx}, $f(\mathring{R})$ theory~, General Relativity, $f(T)$ theory~\cite{Izumi:2012qj}, $f(\phi,T)$ theory~\cite{Gonzalez-Espinoza:2021mwr}. and an action with only teleparallel terms.}
    \label{tab:Scalar_Cases}
\end{sidewaystable}

\normalsize
\subsection{Pseudoscalar Perturbations}

The gauge invariant pseudoscalar $\sigma$ can be treated separately as
\begin{align}
    \udt{e}{A}{\mu} \rightarrow \begin{bmatrix}
    1 & 0\\
    0 & \delta^{Ii}(\delta_{ij} + \epsilon\, \varepsilon_{ijk} \partial^{k}\sigma)
    \end{bmatrix}\,,
\end{align}
The action in Fourier space is expanded up to second order
\begin{align}
    \mathcal{S}_{\text{PS}}^{(2)} &=- \int dt \frac{d^{3}k}{(2\pi)^{3}} \left[ \left(k^{2} \frac{d}{dt}(\dot{\phi} G_{\text{Tele},I_{2}})+ \frac{4}{9} k^{4} \left(G_{\text{Tele},T_{\text{ax}}}-2X G_{\text{Tele},J_{1}}\right)\right) \sigma^{2} \right.\nonumber \\
    &  \qquad \qquad \qquad \qquad \left. + k^{2}\left( \frac{4}{9} G_{\text{Tele},T_{\text{ax}}}+X \left(G_{\text{Tele},J_{5}} + \frac{4}{3} G_{\text{Tele},J_{10}} \right)\right)\dot{\sigma}^{2}\right]\,,
\end{align}
where the first term exhibits higher-order spatial derivatives and the second term exhibits higher order mixed derivatives. For ghost stability,
\begin{align}
    \mathcal{M}_{\text{PS}} = k^{2}\left( \frac{4}{9} G_{\text{Tele},T_{\text{ax}}}+X \left(G_{\text{Tele},J_{5}} + \frac{4}{3} G_{\text{Tele},J_{10}} \right)\right) >0\,.
\end{align}
By performing Fourier transformation for the temporal terms, determinant of the equation is set to zero
\begin{align}
    \omega^{2} \left( \frac{4}{9} G_{\text{Tele},T_{\text{ax}}}+X \left(G_{\text{Tele},J_{5}} + \frac{4}{3} G_{\text{Tele},J_{10}} \right)\right) -  \frac{d}{dt}(\dot{\phi} G_{\text{Tele},I_{2}}) - \frac{4}{9} k^{2} \left(G_{\text{Tele},T_{\text{ax}}}-2X G_{\text{Tele},J_{1}}\right) = 0\,.
\end{align}
When considering the dispersion solution of $\omega^{2} = k^{2} c_{\text{PS}}^{2}$, the propogating speed for the pseudoscalar is given by
\begin{align}
    c_{\text{PS}}^{2} = \frac{ \frac{4}{9} \left(G_{\text{Tele},T_{\text{ax}}}-2X G_{\text{Tele},J_{1}}\right)}{\frac{4}{9} G_{\text{Tele},T_{\text{ax}}}+X \left(G_{\text{Tele},J_{5}} + \frac{4}{3} G_{\text{Tele},J_{10}}\right)} > 0\,,
\end{align}
such that the Laplacian instability can be avoided for 
\begin{align}
    \mathcal{N}_{\text{PS}} =  \frac{4}{9} \left(G_{\text{Tele},T_{\text{ax}}}-2X G_{\text{Tele},J_{1}}\right) > 0\,.
\end{align}
As shown in this section, the action corresponding to the pseudoscalar is dependent on the teleparallel portion of BDLS theory. In fact, for theories such as Horndeski~\cite{Kobayashi:2019hrl_HorndeskiReview}, $f(\phi,T)$~\cite{Gonzalez-Espinoza:2021mwr}, $f(T)$~\cite{Izumi:2012qj}, GBD~\cite{DeFelice:2010jn_Generalised-Brans-Dicke}, Brans-Dicke~\cite{Brans:1961sx}, $f(\lc{R})$~\cite{DeFelice:2010aj-f(R)_theories} and GR, the pseudoscalar reduces to a non-dynamical mode and thus it is a non-propagating mode. This shows that, in Minkowski background, only a few additional scalar invariants, provided by the teleparallel analogue of Horndeski theory, contribute to any additional propagating DoF.

\section{Discussion and Conclusions} \label{sec:conclu}

We have investigated  constraints emerging by considering the stability of perturbations about a Minkowski background in teleparallel analogue of Horndeski gravity. The approach is performed by exploring the  ghost instabilities through looking at potentially negative expressions of  kinetic term  associated with propagating DoFs, as well as Laplacian instabilities  emerging when propagation speeds of perturbations are not positive. This leads to possible exponential growth rates. These considerations are fundamental for the construction of robust and self-consistent cosmological models related to the scalar-tensor gravity.

Specifically, we have discussed  teleparallel analogue of Horndeski gravity where the most general scalar-tensor action is considered, provided that the Lagrangian terms are not parity violating and that scalars are produced by no more than quadratic contractions of  torsion scalar. The naturally lower order nature of teleparallel gravity induces a much richer landscape of functional models when compared with standard metric Horndeski gravity, based on  Levi-Civita connection. In other words, the replacement of torsion with curvature to lead dynamics turns out to manifest as a generalization of the standard Horndeski theory. Importantly, this produces a generalization of  tensor mode propagation speed which means that the restrictions  appearing in standard Horndeski gravity, due to the constraints on the   gravitational waves speed, do not have the same effect here.

Our calculations have  proceeded by first deriving the background equations of motion~(\ref{background-equations-1}--\ref{background-equations-3}) which are obtained by considering a flat FLRW background in which we take the Minkowski limit. We find this to be a convenient approach to determine the system of equations. These are useful equations for reducing the expressions of  perturbations that ensue. The procedure is  described in Sec.~\ref{sec:perturbations}. We then proceeded to directly determine constraints to prevent ghost or Laplacian instabilities starting with  tensor modes which results in the action in Eq.~\eqref{eq:tensor_action} and tensor propagation speed in Eq.~\eqref{eq:tensor_total_speed}. We collected the constraints for popular subclasses of BDLS theory in Table~\ref{tab:Tensor_Cases} where the known results are obtained for standard Horndeski gravity as well as $f(\lc{R})$ and Brans-Dicke gravity, but new conditions are found in other cases. As for the vector perturbations, Sec.~\ref{sec:vector_pert} shows that vector modes are indeed propagating for some cases of  $G_{\rm Tele}$ contribution, namely when $J_i$ scalars appear in  this function. However, these contributions may be small and within observational constraints.

The scalar perturbations contain the most diverse range of DoFs. This rich structure produces the intricate action in Eq.~\eqref{eq:scalar_pert_action}, together with the propagation speed in Eq.~\eqref{eq:scalar_DoF_speed}. As in the case of  tensor modes, constraints for each subclass of popular models is reported in Table~\ref{tab:Scalar_Cases} where  results for the Minkowski perturbations are recovered for standard metric Horndeski gravity together with $f(\lc{R})$ and Brans-Dicke gravity. Interestingly, either no constraints are set on some of the purely teleparallel models or only lightly limited cases are set. Importantly, these constraints are consistent with the tensor mode constraints on the functional models.

These results are promising in terms of the viability of BDLS theory. As future work, it is intriguing to explore  constraints from tachyonic instabilities. These constraints   have been connected to Jeans instability where exponential growth of perturbations is slowed by the expansion of the Universe and where the matter dispersion relation has a negative mass term that renders it to vanish. This could lead to a more general consideration of  perturbations about a flat FLRW background spacetime which would then be directly linked to observations such as those related  to the  growth of large scale structure.
In forthcoming studies, possible observational constraints will be scrutinized.

\acknowledgments{
This paper is based upon work from COST Action CA21136 {\it Addressing observational tensions in cosmology with systematics and fundamental physics} (CosmoVerse) supported by COST (European Cooperation in Science and Technology). SC  acknowledges the  {\it Istituto Nazionale di Fisica Nucleare}  ({\it iniziative specifiche} QGSKY and MOONLIGHT2). MC  acknowledges funding by
the Tertiary Education Scholarship Scheme (TESS, Malta).}

\bibliographystyle{utphys}
\bibliography{references}

\end{document}